\documentclass{article}

\usepackage{arxiv}

\usepackage[utf8]{inputenc} 
\usepackage[T1]{fontenc}    
\usepackage{hyperref}       
\usepackage{url}            
\usepackage{booktabs}       
\usepackage{amsfonts,amsmath}       
\usepackage{nicefrac}       
\usepackage{microtype}      
\usepackage{lipsum}		
\usepackage{graphicx}
\usepackage[numbers]{natbib}
\usepackage{doi}

\usepackage{caption}
\usepackage{subcaption}
\usepackage{graphicx}
\usepackage[table,xcdraw]{xcolor}
\usepackage{float}
\usepackage{multirow}

\usepackage{ulem}

\title{A stochastic least action principle applied in the description of black swan events}

\author{{Tatiana Cardoso e Bufalo} \\
	Department of Physics\\
	Federal University of Lavras\\
		Lavras, MG, Brazil \\
	\texttt{taticardoso@ufla.br} \\
	\And
	{R. Bufalo} \\
	Department of Physics\\
	Federal University of Lavras\\
		Lavras, MG, Brazil \\
	\texttt{rodrigo.bufalo@ufla.br } \\
	\And
	{Lucas P. G. de Figueiredo} \\
	Department of Physics\\
	Federal University of Lavras\\
	Lavras, MG, Brazil \\
\texttt{lucas.figueiredo@estudante.ufla.br } \\
	\And
	{Qiuping A. Wang} \\
	Laboratoire SCIQ, EISEA Numérique et Société,\\ 
	ESIEA Group\\
		Paris, France \\
	Laboratoire IMMM, CNRS\\
	Le Mans University\\
		Le Mans, France\\
	\texttt{alexandre.wang@univ-lemans.fr} \\
	\And
	{Fabio Lucio Alves} \\
	Nanjing University\\
	Nanjing, China \\
	\texttt{fabio.lucio@cern.ch} \\}

\hypersetup{
pdftitle={A stochastic least action principle applied in the description of black swan events},
pdfsubject={Physics},
pdfauthor={Lucas Figueiredo},
pdfkeywords={Stochastic processes; Black swan events; Stable distributions},
}

\begin{document}
\maketitle

\begin{abstract}
In this paper we present a formulation of the stochastic least action principle (SAP) to encompass random movements with black swan events (of non dissipative systems) in terms of heavy tailed distributions.
The black swan events are rare and drastic events, such as earthquakes and financial crisis.
It has been observed that the Tsallis entropy suits well in the description of the black swan events rather than the Shannon-Boltzman-Gibbs entropy, which is intrinsically related to the fact that black swan events of physical systems are proportional to non-local correlations. 
As a consequence, we could assess the validity of the path probability distribution obtained using the non-additive Tsallis entropy.
\end{abstract}


\section{Introduction}

The least action principle (LAP), also known as Hamilton’s principle, is one of the most important principles in physics. It is related to a fundamental rule of nature that a system evolves in such a way to extremize an energy related quantity called action \cite{gold}. It is known that LAP applies only to Hamiltonian systems having energy conservation in deterministic motion \cite{gold}. More recently, LAP has been extended to a random motion of Hamiltonian systems which statistically conserve energy (see ref. ~\cite{lin} and references therein for a review), a type of stochastic process that can be used to describe diffusion \cite{demichev}. This extension has been called stochastic action principle (SAP), which is motivated by the century old conflict between the second law of thermodynamics and the Liouville's theorem in the classical mechanics, as discussed in the refs. \cite{wang1,wang2}, as well as by the question whether and how the path probability depends on the action in random motion of non-dissipative systems \cite{lin2012}. The predictions of SAP has been verified by the outcomes of the numerical experiments of random motion with a Gaussian noise \cite{lin, lin2012}, which confirm, from a probabilistic point of view, the existence of a classical analogue of the Feynman factor $e^{iA}$ for the path integral formalism \cite{feynman2010, kleinert}. This factor also unveiled a close relationship between the path integral formalism and the action functional A in classical mechanics. 


This work aims to generalize the previous mentioned finding to a different class of motion, those not described by a Gaussian distribution, since the SAP formulation is not related to any specific type of random motion.
In particular, we shall focus our attention on examining the validity of the SAP in the description of black swan events, a type of rare and extreme event \cite{taleb}.
In order to carry out this analysis, we shall consider random noises generated by a stable distribution, a distribution which can be heavy tailed depending on the choice of its free parameter (see equation \eqref{stable}).
Essentially, this modification, corresponds to an anomalous diffusion processes, presenting long range interactions \cite{klages}.
A natural framework to describe this kind of non-local (long range) correlation is the Tsallis $q$-entropy \cite{tsallis_orig}.

The Tsallis $q$-entropy was formulated in the context of multifractals and was initially presented aiming the possibility of generalizing the Shannon-Boltzmann-Gibbs entropy \cite{tsallis_orig}.
Among the many interesting properties that the use of the Tsallis entropy presents, one can observe that it describes systems with long range interactions, while the Shannon-Boltzmann-Gibbs entropy is limited to systems with short-range interactions \cite{landsberg, abe}.
This last statement enunciates that the Tsallis $q$-entropy is the legitimate framework for the description of anomalous diffusion processes \cite{tsallisdiff}.
Moreover, the Tsallis $q$-entropy has been applied to a wide and vast class of several physical systems, ranging from  solar winds measurements, chaos, financial markets, quantum information, high energy physics and even to black hole entropy analysis \cite{tsallis_gell, tsallis2021, caos, xiao2021innovation, majhi2017non}.

Another important point worth to mention about the Tsallis $q$-entropy is that for suitable values of the parameter $q$, one expects that small probabilities become more significant in the characterization of the system.
As can be noticed, that is an essential feature to consider in the description of black swan events.
Hence, the properties discussed above strongly support the use of the Tsallis entropy for the description of the black swan events within the SAP formalism.

In this paper we present an analysis of the black swan events (of non dissipative systems) within the stochastic least action principle, supplemented by the Tsallis $q$-entropy in order to take into account the non-local (long range) correlations present in this class of physical system.
In section ~\ref{sec:sap}, we review the main aspects of the stochastic least action principle, as well as the presence of constraints, to obtain a path probability distribution in terms of the maximization of the entropy.
In section ~\ref{sec:simgaus}, we develop the path probability distribution to the random motion, governed by the Gaussian noise, using the Shannon entropy and the Fokker-Planck diffusion equation.
Furthermore, we verify the validity of the results by means of computational simulations for the case of null (free particle), linear (constant force) and quadratic (simple harmonic oscillator - SHO) potentials.
The section ~\ref{sec:black} presents the main analysis of this work, where we develop the SAP for the case of black swan events within the Tsallis $q$-entropy in terms of stable distributions.
The results and perspectives are summarized in the section ~\ref{sec:conc}.


\section{The stochastic least action principle}
\label{sec:sap}

The classical least action principle states that the action path must be minimal: $\delta A=0$.
However, in a scholastic context, which is probabilistic, we have to state analogously that \cite{lin}
\begin{equation}
\overline{\delta A}=\sum_{k}p_{k}\delta A_{k}=0,
\label{estat}
\end{equation}
where $p_k$ is the probability distribution chosen to describe the physical system.
Moreover, using the mean action of the paths, i.e., $\overline{A}=\sum \limits_{k}p_{k}A_{k}$, and performing a variation of it, we find
\begin{equation}
\delta\overline{A}=\delta Q+\overline{\delta A}.
\label{variat}
\end{equation}
The quantity $\delta Q$ is identified as the varentropy of the system.
This notion comes from the thermodynamics, in which the quantities of a isolated system as energy ($E$), temperature ($T$) and entropy ($S$) satisfy the following equation of state
\begin{equation} \label{eos}
T\delta S=\delta E=\sum_{i}\delta p_{i}E_{i}=\delta\overline{E}-\overline{\delta E}.  
\end{equation}
Hence, by comparing the equations \eqref{variat} and \eqref{eos}, we can establish that
\begin{equation}
\delta Q=\frac{\delta S}{\gamma}.
\label{varent}
\end{equation}
Finally, replacing the equations \eqref{estat} and \eqref{varent} into equation \eqref{variat}, it results in
\begin{equation}
\delta\left(S-\gamma\sum_{k}p_{k}A_{k}+\alpha\sum_{k}p_{k}\right)=0,
\label{const}
\end{equation}
where we have added the {\it natural constraint}, $\sum\limits_{k}p_{k}=1$, by means of obtaining a normalized probability distribution.
In the expression \eqref{const}, the quantities $\gamma$ and $\alpha$ are interpreted as Lagrange multipliers.
Therefore, we can conclude that, in order to obtain a path probability distribution in terms of the stochastic least action principle, we need to maximize the entropy $S$ subjected to the constraints in \eqref{const}.
In other words, this statement is the Jaynes maximum entropy principle \cite{jaynes}.


\section{The path probability distribution of random motion with Gaussian noise}
\label{sec:simgaus}

One of the most important physical formulation of entropy, which allowed to establish a relation between the statistical mechanics and the thermodynamics, is the Shannon-Boltzmann-Gibbs entropy \cite{reif}:
\begin{equation}
    S=-\sum_{k}p_{k}\ln p_{k}.
    \label{shannon}
\end{equation}
Its physical applications are very abundant, as the information theory \cite{shannonteo},  hydrology \cite{entropy_hidro}, image processing \cite{imag_entropy}, quantum information \cite{entropy_quant} and so on.
We can maximize the entropy equation in \eqref{shannon} using the Lagrange formalism.
Thus, considering the constraints \eqref{const}, we can write the following Lagrangian
\begin{equation}
    \mathcal{L}=-\sum_{k}p_{k}\ln p_{k}-\gamma\left(\sum_{k}p_{k}A_{k}-\langle A\rangle\right)+(1-\alpha)\left(\sum_{k}p_{k}-1\right).
\end{equation}
We can evaluate the Euler-Lagrange equation of motion as
\begin{equation}
    -\ln p_{i}-\alpha-\gamma\langle A\rangle=0.
    \label{this}
\end{equation}
The equation \eqref{this} can be reorganized using the constraints equations, resulting in
\begin{equation}
    p_{k}=\frac{1}{Z}e^{-\gamma A_{k}},
    \label{path_prob}
\end{equation}
where $Z=\sum\limits _{i}e^{-\gamma A_{i}}$ is the normalization factor, which corresponds to a distribution with the shape of the Boltzmann-Gibbs distribution.

In order to examine the physical content of the equation \eqref{path_prob}, we perform computational simulations using the Langevin description of random motion of particles\footnote{A type of mechanical stochastic processes where we can consider the trajectories as markovian chains \cite{demichev} and based on that compute its path probability.}, a microscopical approach of diffusion processes \cite{demichev}. This formalism is based on the solutions of the following equation of motion
\begin{equation}
    m\frac{d^{2}x}{dt^{2}}=-\frac{dV(x)}{dx} - m\zeta\frac{dx}{dt} - R,
    \label{langevin_differ}
\end{equation}
where $m$ is the mass of the particles, $x(t)$ is the position, $t$ is the time, $V$ is the potential, $\zeta$ is the dissipation constant and $R$ is a random force.
We consider a non-dissipative scenario ($\zeta \rightarrow 0$) where the system can be described by the conservative forces and the Langevin equation becomes the Newtonian second law of motion with a noise superposition.
The Langevin approach is equivalent to the macroscopic description using the diffusion equation
\begin{equation}
    \frac{\partial p}{\partial t}-D \frac{\partial^{2}p}{\partial x^{2}}=0,
    \label{diff_eq}
\end{equation}
also known as the Fokker-Planck equation, where $p(x)$ is the position probability.
The solutions of the equation \eqref{langevin_differ},can be written as a superposition of the classical path $f$ with a random noise $\chi$, resulting in
\begin{equation}
x_{i}=x_{i-1}+\chi_{i}+f(t_{i})-f(t_{i-1}).
    \label{trajectory}
\end{equation}
The explicit expression of the noise $\chi$ is defined in terms of the type of the diffusion process we are observing.
The most known diffusion process in the literature is the Brownian motion \cite{browniano}, the normal diffusion process characterized by the Gaussian distribution, i.e., $\chi$ is a random noise gaussianally distributed.

\begin{figure}[h]
     \centering
     \begin{subfigure}[b]{0.33\textwidth}
         \centering
         \includegraphics[width=\textwidth]{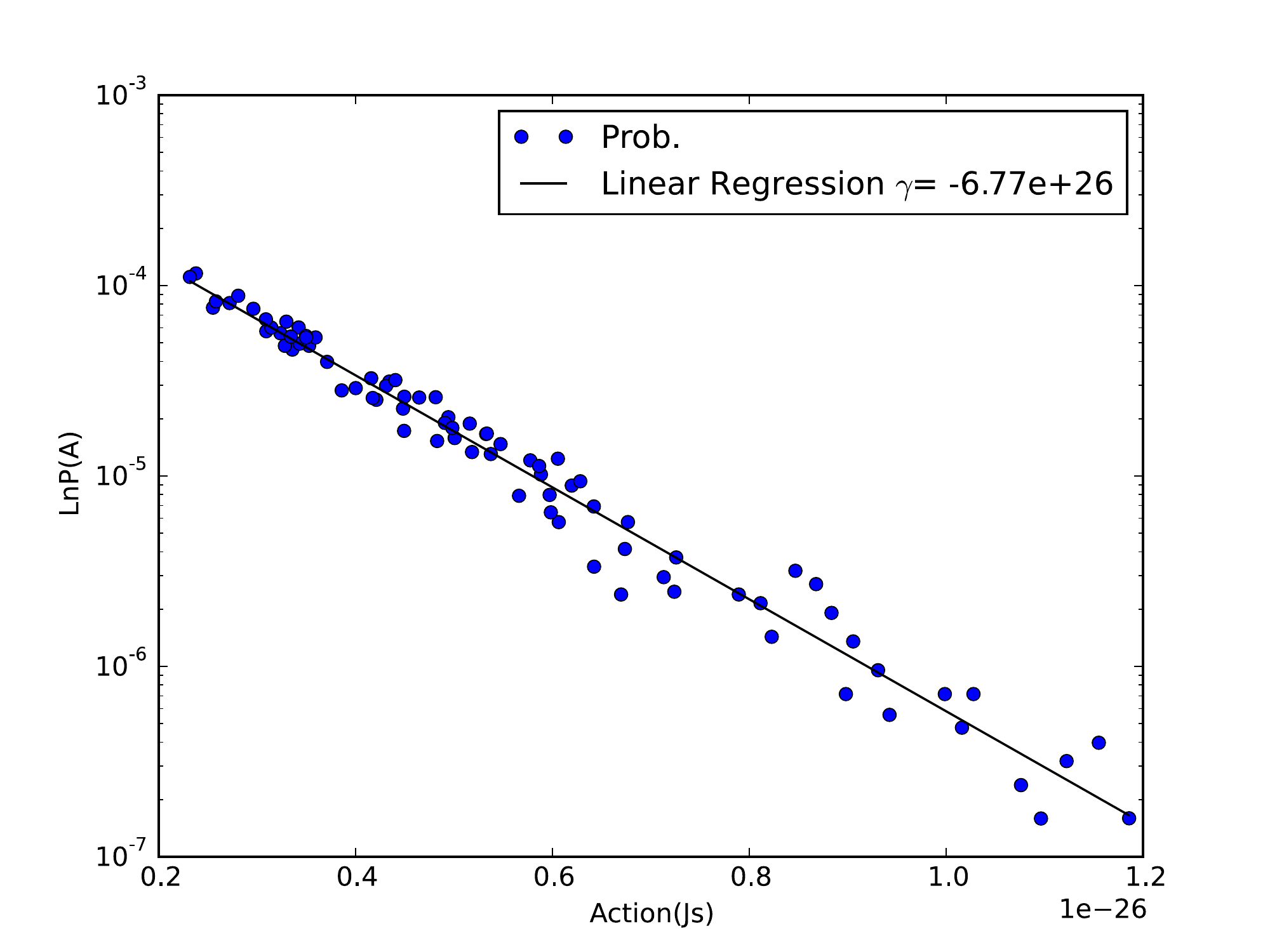}
         \caption{ }
     \end{subfigure}
     \begin{subfigure}[b]{0.33\textwidth}
         \centering
         \includegraphics[width=\textwidth]{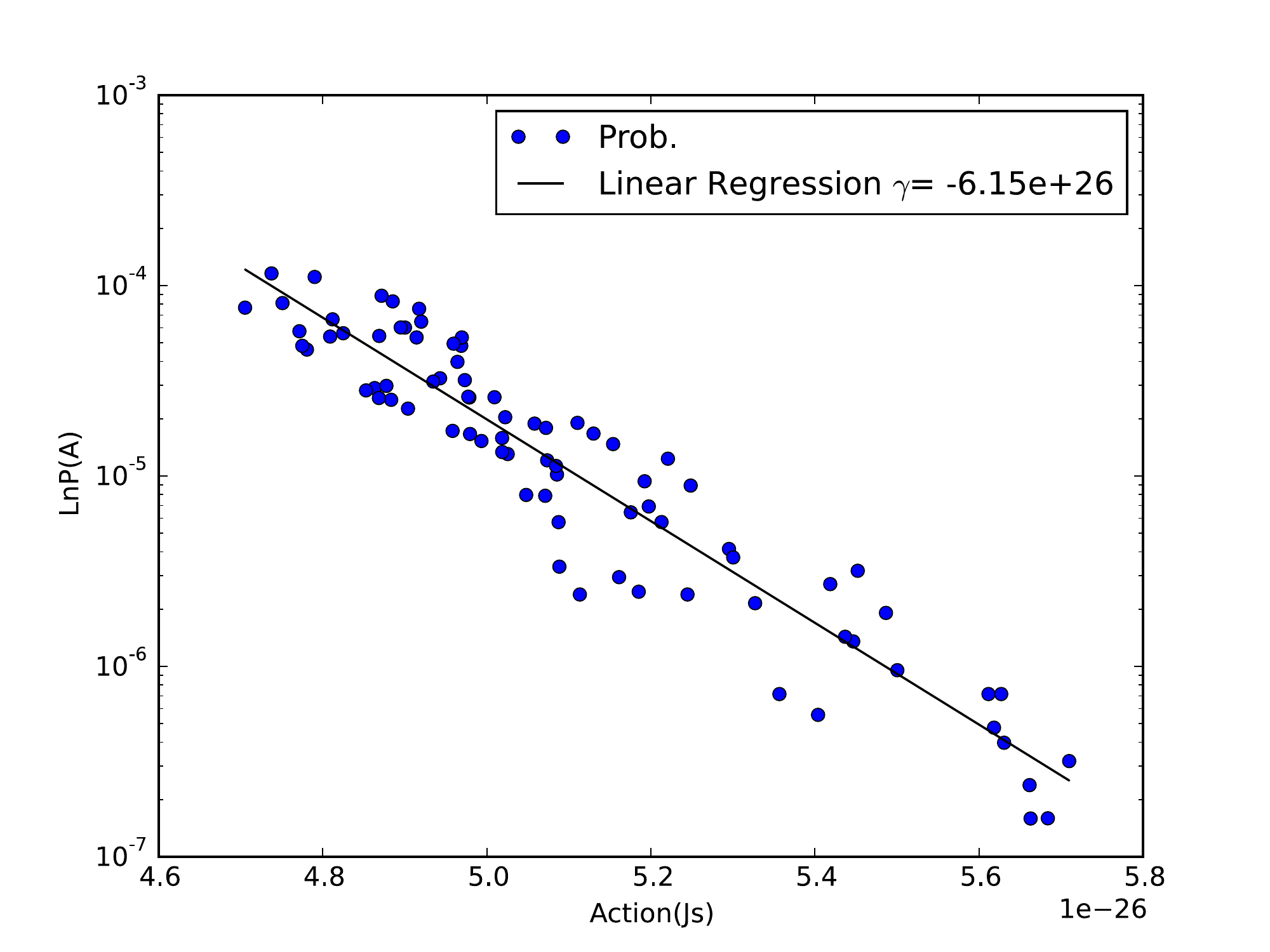}
                  \caption{ }
     \end{subfigure}
     \begin{subfigure}[b]{0.33\textwidth}
         \centering
         \includegraphics[width=\textwidth]{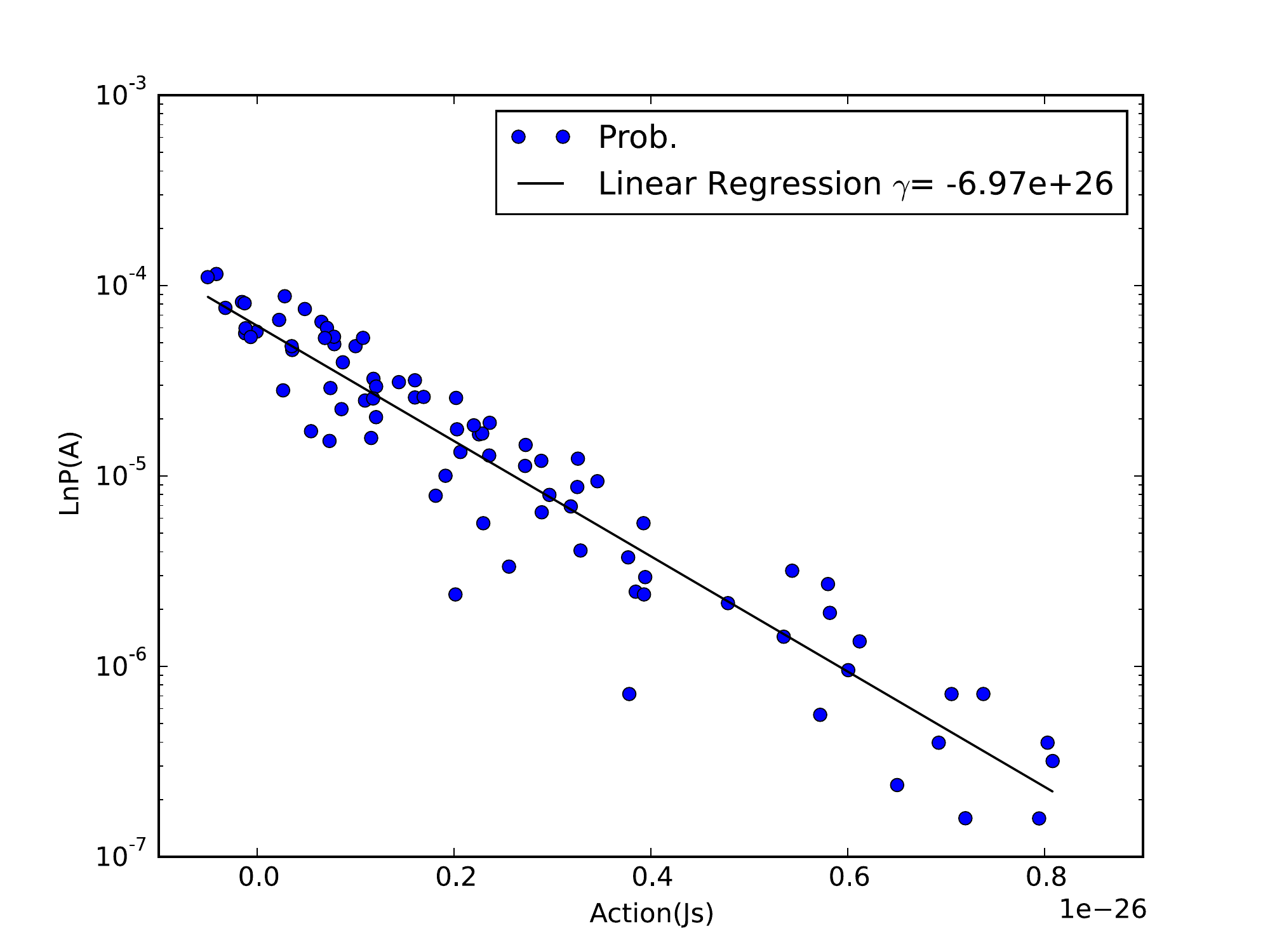}
                 \caption{ }
     \end{subfigure}
        \caption{Simulations of the path probability, logarithm of the probability versus the action of the paths, with Gaussian noise for three different scenarios: (a) null potential, (b) linear potential and (c) quadratic potential.}
        \label{fig:gaus_sim}
\end{figure}

The simulation is prepared considering the following parameters: the trajectories are discrete with $10$ time steps with size $dt=10^{-5}s$, the particles are identical with mass $m=1.39 \times 10^{-15} kg$ and the Gaussian distribution has standard deviation $\sigma=\sqrt{2Ddt} \approx 3\times10^{-9}m$, where $D=4.3\times10^{-13}m^{2}/s$ is the diffusion constant.
The chosen values of these parameters correspond to the scenario of particles of the molecule $SiO_{2}$ in water at normal temperature and pressure conditions, as suggested in ref.~\cite{lin}.  
Under these considerations, we create $10^{3}$ trajectories with width $\delta=\sigma$, launching $10^{6}$ particles for $10^{2}$ times.
The action of the paths is written in terms of a discrete approximation as
\begin{equation}
A_{k}=\sum_{i=1}^{10}\left[\frac{1}{2}m\left(\frac{x_{i+1}-x_{i}}{dt}\right)^{2}-V\left(\frac{x_{i}+x_{i+1}}{2}\right)\right]dt.    
\end{equation}
We can understand the probability of each path as the ratio between the number of particles that moved along each trajectory and the number of particles that arrive at the final point.
As we can observe in figure \ref{fig:gaus_sim}, the paths' probabilities behave as predicted by the expression \eqref{path_prob} for different potentials: null ($V=0$), linear ($V=mgx$, with $g=10m/s^{2}$) and quadratic ($V=m \omega^{2} x^{2}/2$, with $\omega=2,5\times10^{4}rad/s$ \cite{lin}).

Furthermore, for the systems analyzed in the simulations, as depicted in figure \ref{fig:gaus_sim}, we observe that the $\ln P(A)$ and the $A$ are strongly correlated, verifying therefore the exponential probability distribution in equation \eqref{path_prob}.
The values of the correlation and the parameter $\gamma$ (Lagrange multipler) are presented in the table \ref{tab:table_gauss}.
In addition, it was found that by using an equivalent approach to calculate the path probability to the normal diffusion process, known as the Wiener path integral \cite{mazo}, that  $\gamma=\frac{1}{2mD}=-8.36 \times 10^{26} J^{-1}s^{-1}$ \cite{wang1}, which is in good agreement with the simulation results, see table \ref{tab:table_gauss}.
If we think in a analogous way as in the thermodynamics where we have $\gamma \propto T^{-1}$ \cite{reif}, the physical meaning of $\gamma$ in our analysis is the same.
In the thermodynamics, we have that the temperature increases the energy of the states that are distributed by the Boltzmann-Gibbs statistics.
On the other hand, in our ensemble, we observe that the diffusion constant can increase the action of the (allowed) paths of the system. 

\begin{table}[H]
\centering
\begin{tabular}{|
>{\columncolor[HTML]{EFEFEF}}c |c|c|}
\hline
Potential & \cellcolor[HTML]{EFEFEF}Correlation & \cellcolor[HTML]{EFEFEF}$\gamma$ ($J^{-1}s^{-1} $) \\ \hline
Null      & 0.987 & $-6.77 \times 10^{26}$ \\ \hline
Linear    & 0.948 & $-6.15 \times 10^{26}$ \\ \hline
Quadratic & 0.948 & $-6.97 \times 10^{26}$ \\ \hline
\end{tabular}
\caption{Numerical results of the Gaussian simulation.}
\label{tab:table_gauss}
\end{table}

\section{The path probability with black swans events}
\label{sec:black}

\begin{figure}[h]
    \centering
    \includegraphics[scale=0.5]{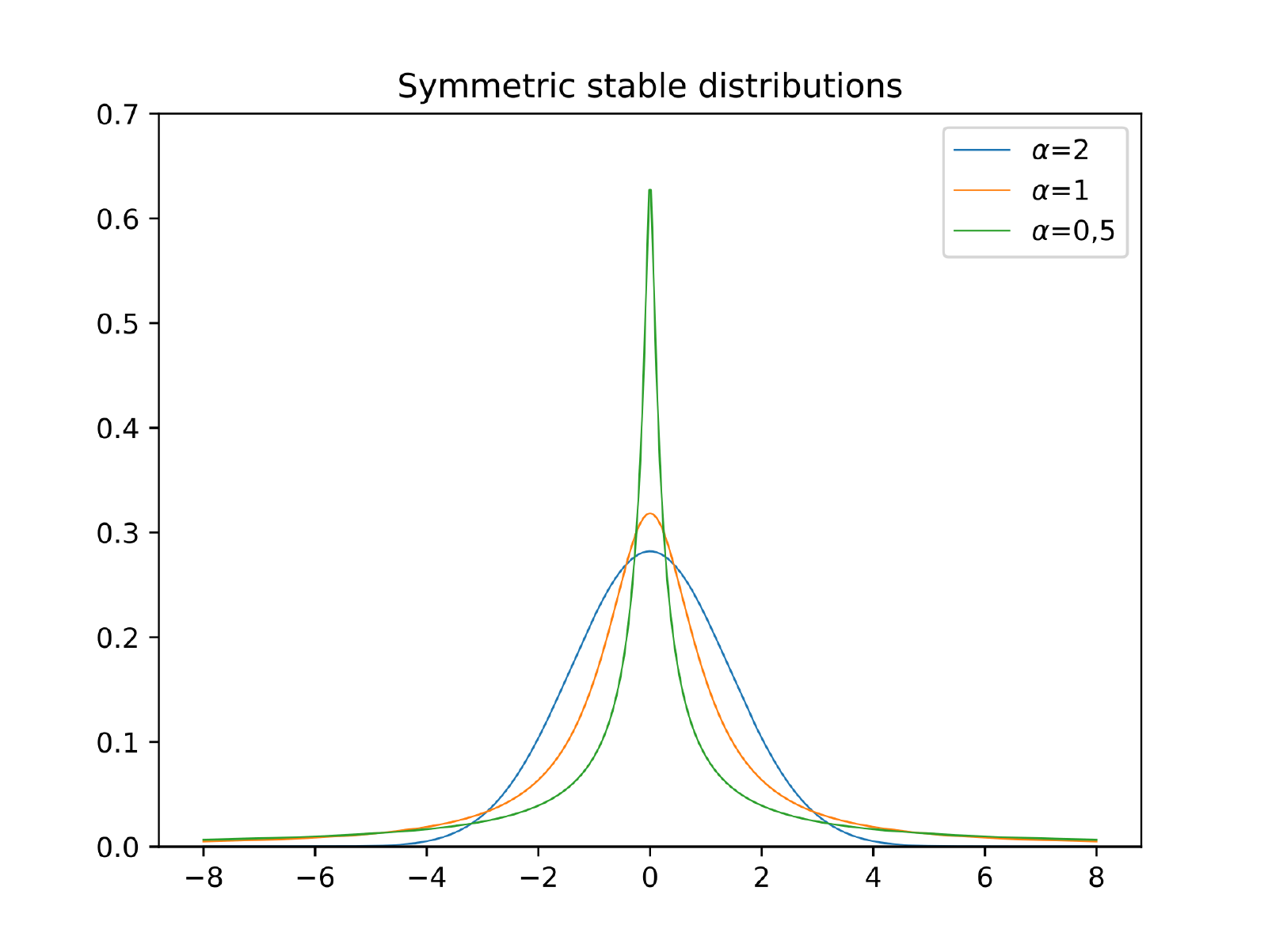}
    \caption{Stable distributions with different values of $\alpha$}
    \label{fig:stable}
\end{figure}

Throughout this section, we are interested to establish the framework of the stochastic least action principle for the case of black swan events.
In order to reach this goal, we will examine the path probability generated for this type of event. 
The proposed analysis consists in modifying the statistical distribution of the random noise $\chi$ in the equation \eqref{trajectory} to a heavy tailed distribution instead of a Gaussian distribution \cite{kleinlevy}. 
This is based on the fact that for heavy tailed distributions, the probability of extreme events are higher as heavier are the tails.
That being said, we choose to work with the symmetric stable distribution due to its close relation with the Gaussian distribution \cite{klages}, as can be seen in figure \ref{fig:stable} for the sake of illustration.
The characteristic function of the stable distribution is
\begin{equation}
    \varphi(k)=e^{-(ck)^{\alpha}},
    \label{stable}
\end{equation}
where the parameter $\alpha$ is the stability parameter, which controls the heaviness of the tails and varies within the interval $(0,2]$, such that the smaller the parameter $\alpha$ is, the heavier are the tails.
Moreover, we are able to recover the Gaussian distribution by considering the following choices: $\alpha = 2$ and $c=\sqrt{\sigma^{2}/2}$, where $\sigma$ corresponds to the standard deviation.
Therefore, we will use the relation between $c$ and $\sigma$, $c=\sqrt{\sigma^{2}/2}$, to obtain the value of the $c$ parameter of the stable distribution \eqref{stable} used on the simulations of black swan events.

Although the use of a heavy tailed symmetric stable distribution to describe black swan events might seems to be naive and straightforward, we observe that, when performing the simulations with this new random noise distribution, the path probability distribution does not have a linear correlation with the action on a log-normal plot.
This can be understood because, besides the change in the random noise probability distribution, we necessarily modify the diffusion behavior of the physical system.
Actually, macroscopically, this modification means that the diffusion process is anomalous and  no longer described by the ordinary diffusion equation \eqref{diff_eq}.
Rather, this diffusion process is governed by a fractional Fokker-Planck equation,
\begin{equation}
    \frac{\partial p}{\partial t}-D_{\alpha} \frac{\partial^{\alpha}p}{\partial x^{\alpha}}=0,
    \label{fokker}
\end{equation}
where $p(x)$ is the position probability distribution (which is the stable distribution), $\alpha$ is the parameter defined in the equation \eqref{stable} and $D_\alpha$ is the related diffusion constant.
Since $\alpha$ assumes values within the interval $(0,2]$, we observe the presence of a fractional differential operator in the spatial derivative in \eqref{fokker}, which is in fact a non-local operator \cite{frac_aplic}.
This implies the presence of non-local interactions in our system, which means that the position probability $p$ is affected not only by its neighborhood, but also by distant points.
This kind of phenomena is observed, for instance, in turbulent diffusion processes \cite{klages,fick_nonlocal}.

Due to the presence of the non-locality, the path distribution analysis is no longer valid since the description developed in sections \ref{sec:sap} and \ref{sec:simgaus} works well to short-range interactions \cite{landsberg}.
This is a well-known condition in the context of the Boltzmann-Gibbs distribution to the thermodynamics, but also a limitation of this formalism.
Owing to this drawback, new approaches for defining entropy have been developed to encompass systems with long-range correlations \cite{abe}.
Within this context, a suitable framework is the Tsallis entropy, also known as $q$-entropy \cite{tsallis_orig,tsallis_gell}.
The main aspect of this proposal is the presence of an $q$ exponent (which characterizes the physical system in consideration) in the probability distribution of the system \cite{tsallis_orig} (the Boltzmann-Gibbs distribution is recovered in the limit $q\to 1$).

Another feature to emphasize about the Tsallis $q$-entropy, extremely important in the description of the black swan events, is that for suitable values of the parameter $q$, small probabilities become more significant and relevant in the description of the system.
Furthermore, one can mention that the Tsallis entropy plays a role in the context of the anomalous diffusion processes in an equivalent way as the Shannon entropy to the normal diffusion \cite{tsallisdiff}.
Therefore, based on the aforementioned properties, one can say that the Tsallis entropy is the natural framework to describe black swan events within the SAP formalism in terms of stable distribution.

The Tsallis entropy can be defined in terms of \cite{tsallis_gell}
\begin{equation}\label{tsallis_entropy}
    S_{q}=-\sum_{i}p_{i}^{q}\ln_{q}p_{i}, 
\end{equation}
where $\ln_{q}x$ is a generalization of the logarithm function
\begin{equation}
    \ln_{q}x=\frac{x^{1-q}-1}{1-q}.
\end{equation}
The original logarithm function, and consequently the Shannon entropy, is recovered in the limit $q\rightarrow1$.
In addition, the inverse of this function is a generalization of the exponential function
\begin{equation}
    e_{q}^{x}=\left[1+(1-q)x\right]^{\frac{1}{1-q}}.
\end{equation}
All the properties of the Shannon-Boltzmann-Gibbs entropy are satisfied by the Tsallis entropy \eqref{tsallis_entropy}, except the property of additivity.
Actually, for two independent systems $A$ and $B$, the Tsallis entropy has a pseudo additivity property
\begin{equation}
    S_{q}(A+B)=S_{q}(A)+S_{q}(B)+(1-q)S_{q}(A)S_{q}(B).
\end{equation}

Hence, in order to obtain the path probability distribution for the random motion encompassing black swan events, within the framework  developed in sections \ref{sec:sap} and \ref{sec:simgaus}, we shall maximize the Tsallis entropy $S_{q}$ equation \eqref{tsallis_entropy} subject to the constraints obtained by the stochastic least action principle \eqref{const}.
Nonetheless, when the Tsallis entropy is considered, the constraints \eqref{const} must be modified to avoid divergences in the partition function (as well as in the energy of the system) \cite{tsallis_vinc}.
Moreover, this modification is also important to the maintenance of the thermodynamic Legendre structure.
Hence, under these considerations, we find the new constraint equation
\begin{equation}\label{const_ts}
    \delta\left(S_{q}-\gamma'\frac{\sum\limits_{k} p_{k}^{q}A_{k}}{\sum\limits_{k} p_{k}^{q}}+\alpha\sum_{k}p_{k}\right)=0.
\end{equation}
Furthermore, considering the equation \eqref{const_ts} and the Lagrange formalism, the path probability distribution can be extracted from the Lagrangian
\begin{equation} 
    \mathcal{L}=\left(\frac{1-\sum\limits_{k} p_{k}^{q}}{q-1}\right)+\alpha\left(\sum_{k}p_{k}-1\right)-\alpha\gamma'(q-1)\left(\sum_{k}p_{k}A_{k}-\langle A\rangle\right).
\end{equation}
Thus, evaluating the Euler-Lagrange equations of motion, we determine the following distribution
\begin{equation}
    p_{k}=\frac{e_q^{-\gamma A_k}}{\sum\limits_i e_q^{-\gamma A_i}}
    \label{path_gen}
\end{equation}
where $\gamma \equiv \frac{\gamma '}{\sum\limits_{k} p_{k}^{q}}$.
We have also performed the replacement $A_{k} \to A_{k}- \langle A \rangle $, being $\langle A \rangle = A_{\mbox{clas}}$, since the noise distribution is symmetric.

\begin{figure}[H]
     \centering
     \begin{subfigure}[b]{0.43\textwidth}
         \centering
         \includegraphics[width=\textwidth]{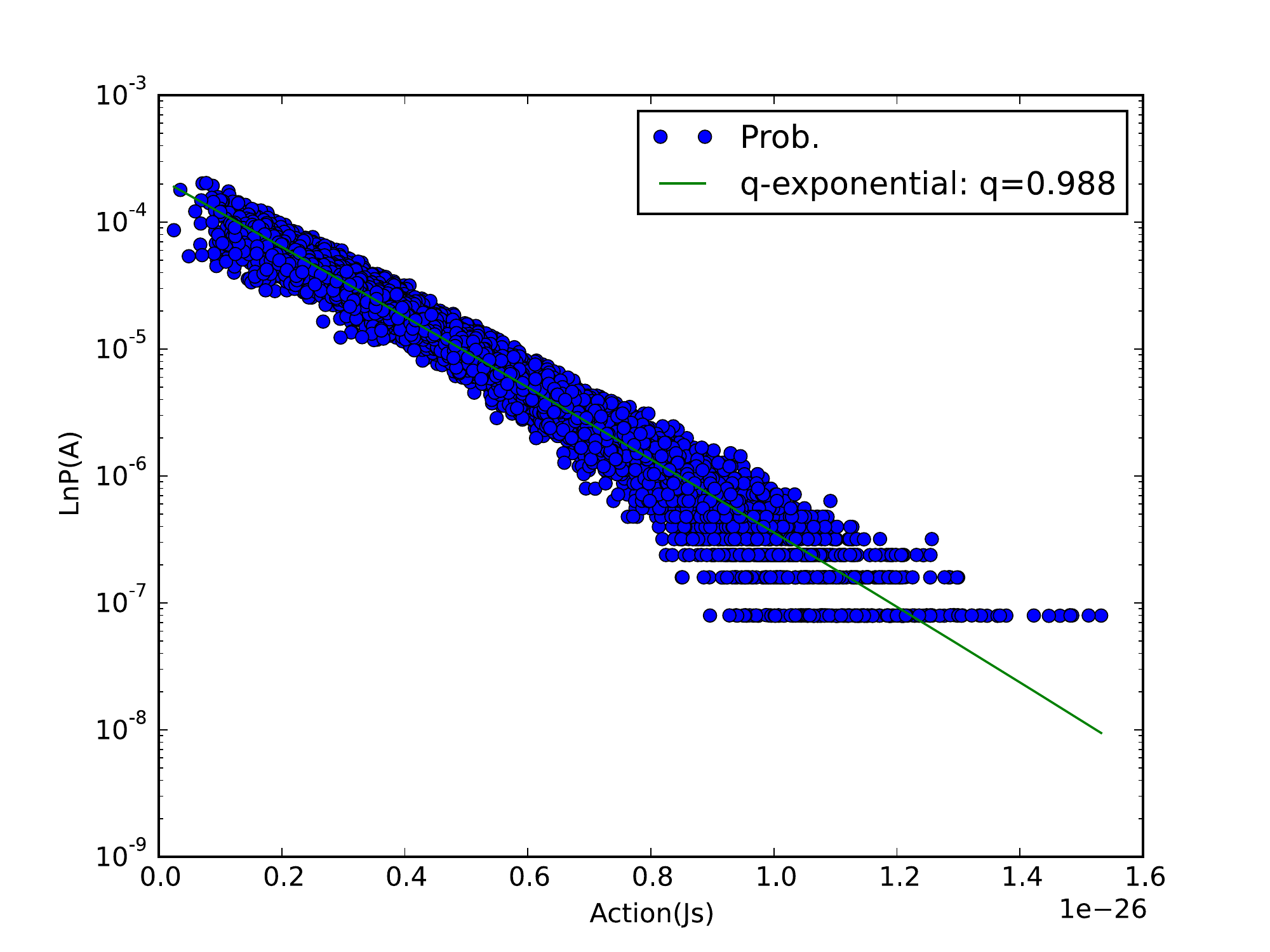}
         \caption{$\alpha=2$}
     \end{subfigure}
     \begin{subfigure}[b]{0.43\textwidth}
         \centering
         \includegraphics[width=\textwidth]{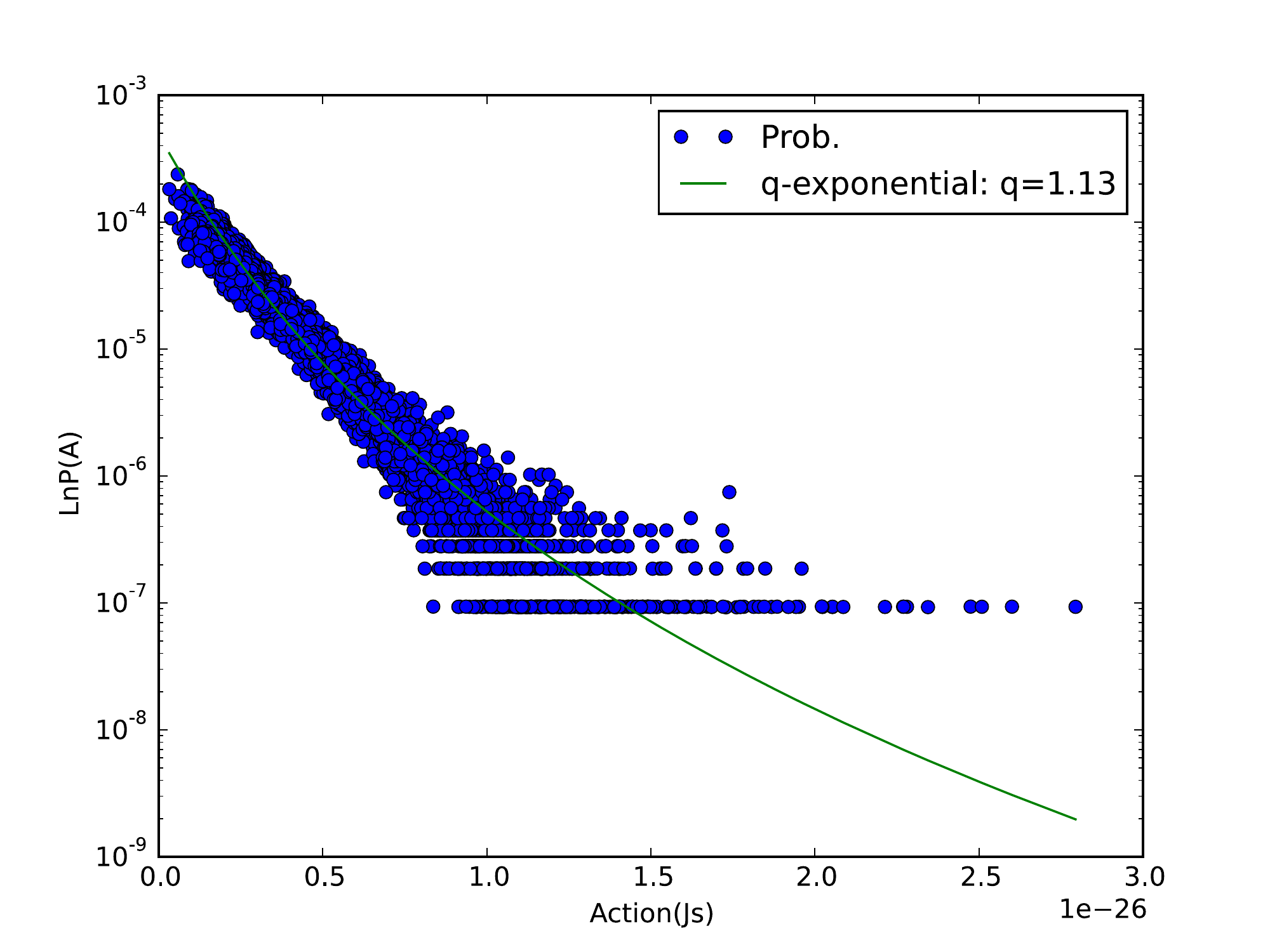}
         \caption{$\alpha=1.75$}
     \end{subfigure}
     \begin{subfigure}[b]{0.43\textwidth}
         \centering
         \includegraphics[width=\textwidth]{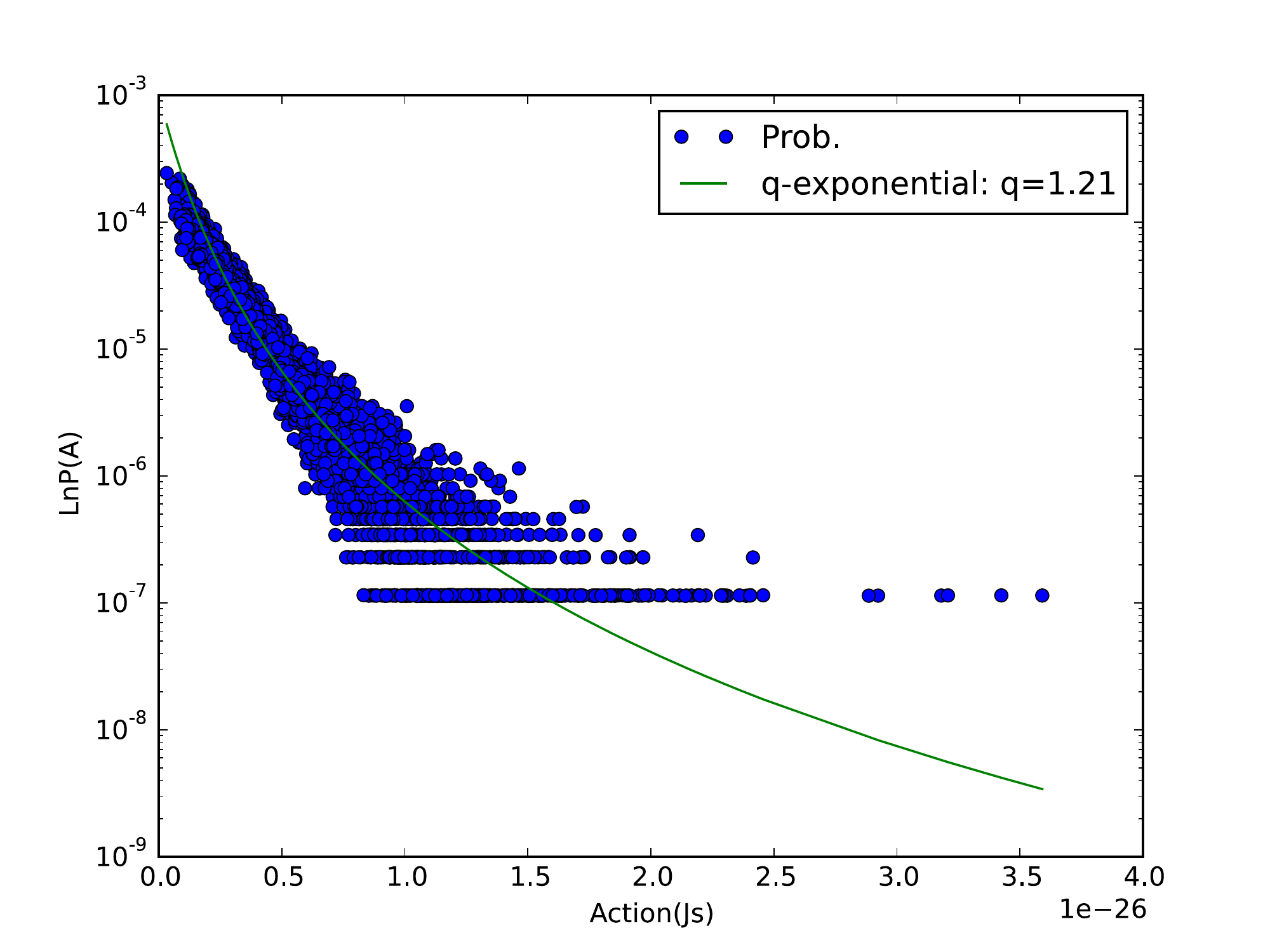}
         \caption{$\alpha=1.5$}
     \end{subfigure}
    \begin{subfigure}[b]{0.43\textwidth}
         \centering
         \includegraphics[width=\textwidth]{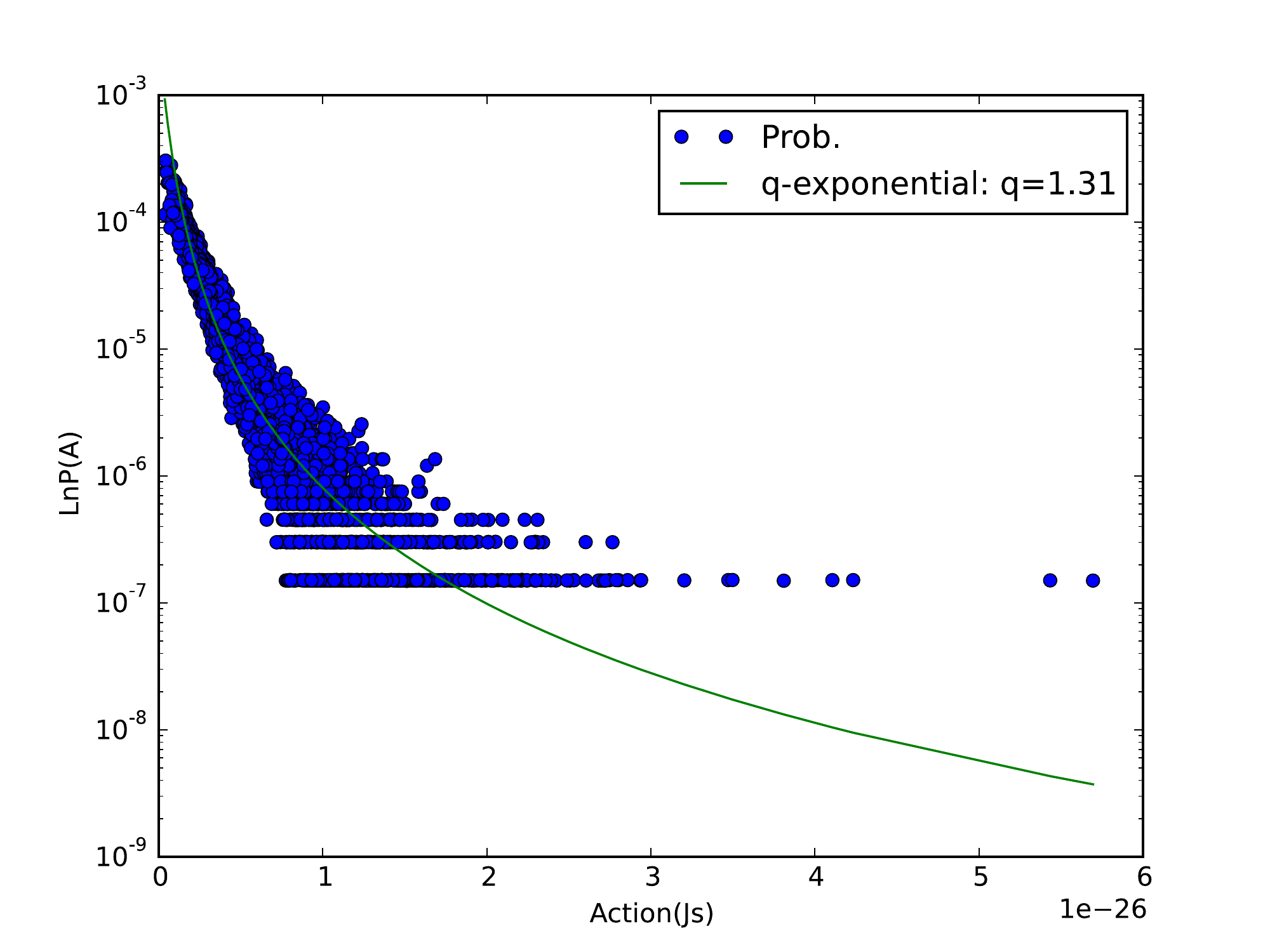}
         \caption{$\alpha=1.25$}
     \end{subfigure}
     \begin{subfigure}[b]{0.43\textwidth}
         \centering
         \includegraphics[width=\textwidth]{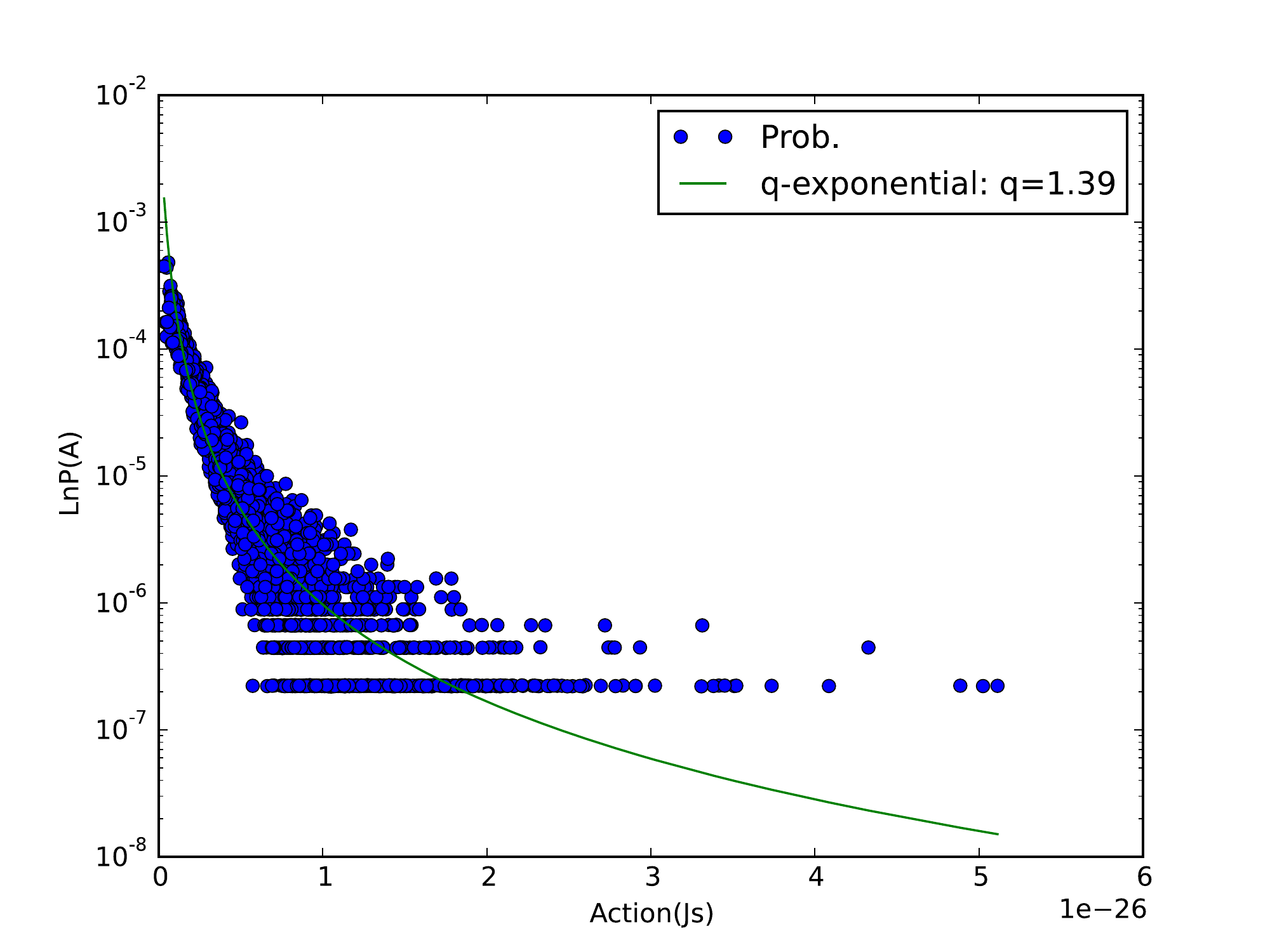}
   \caption{$\alpha=1$}
     \end{subfigure}
     \begin{subfigure}[b]{0.43\textwidth}
         \includegraphics[width=\textwidth]{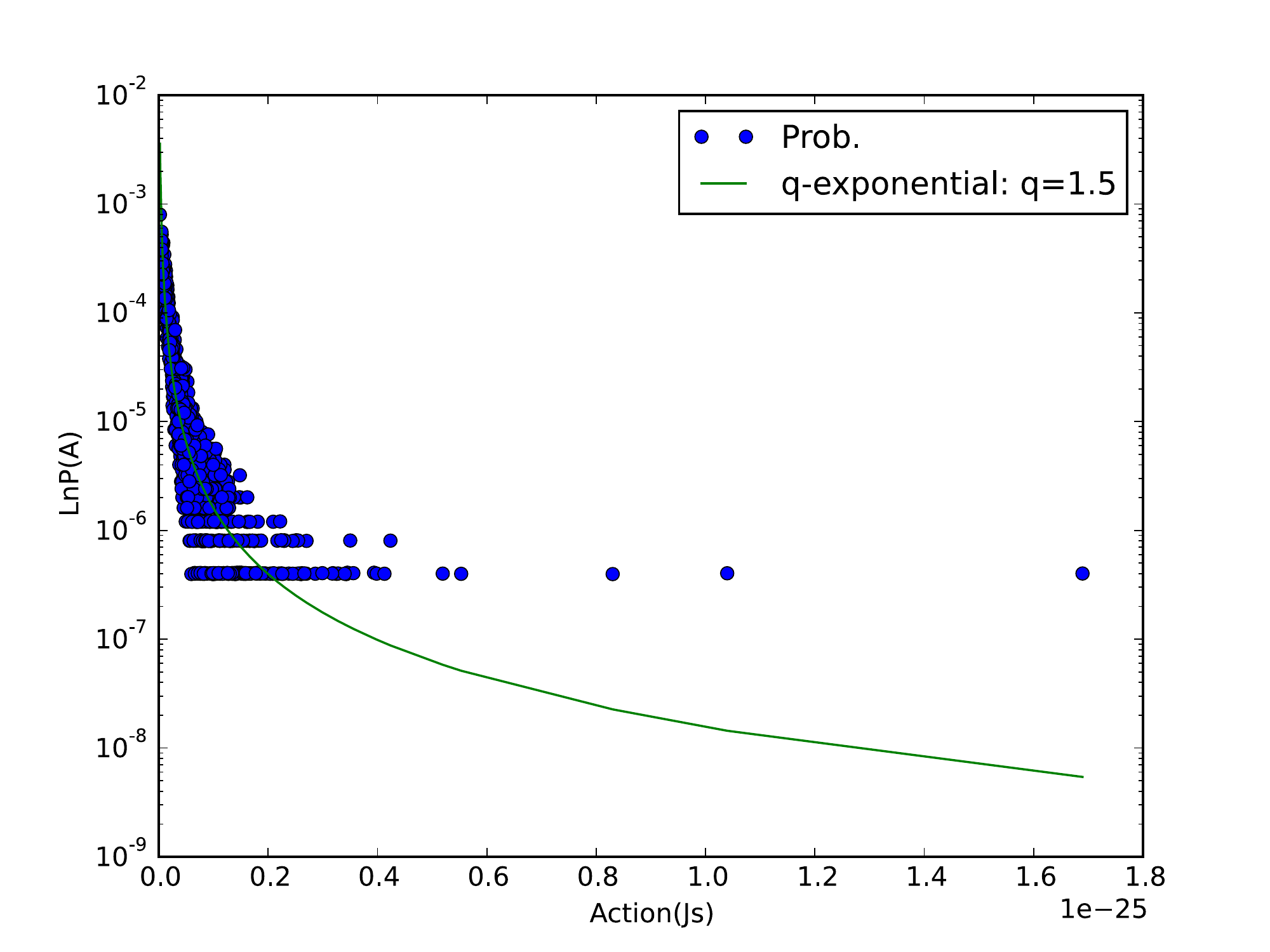}
\caption{$\alpha=0.75$}
     \end{subfigure}
    \begin{subfigure}[b]{0.43\textwidth}
         \centering
         \includegraphics[width=\textwidth]{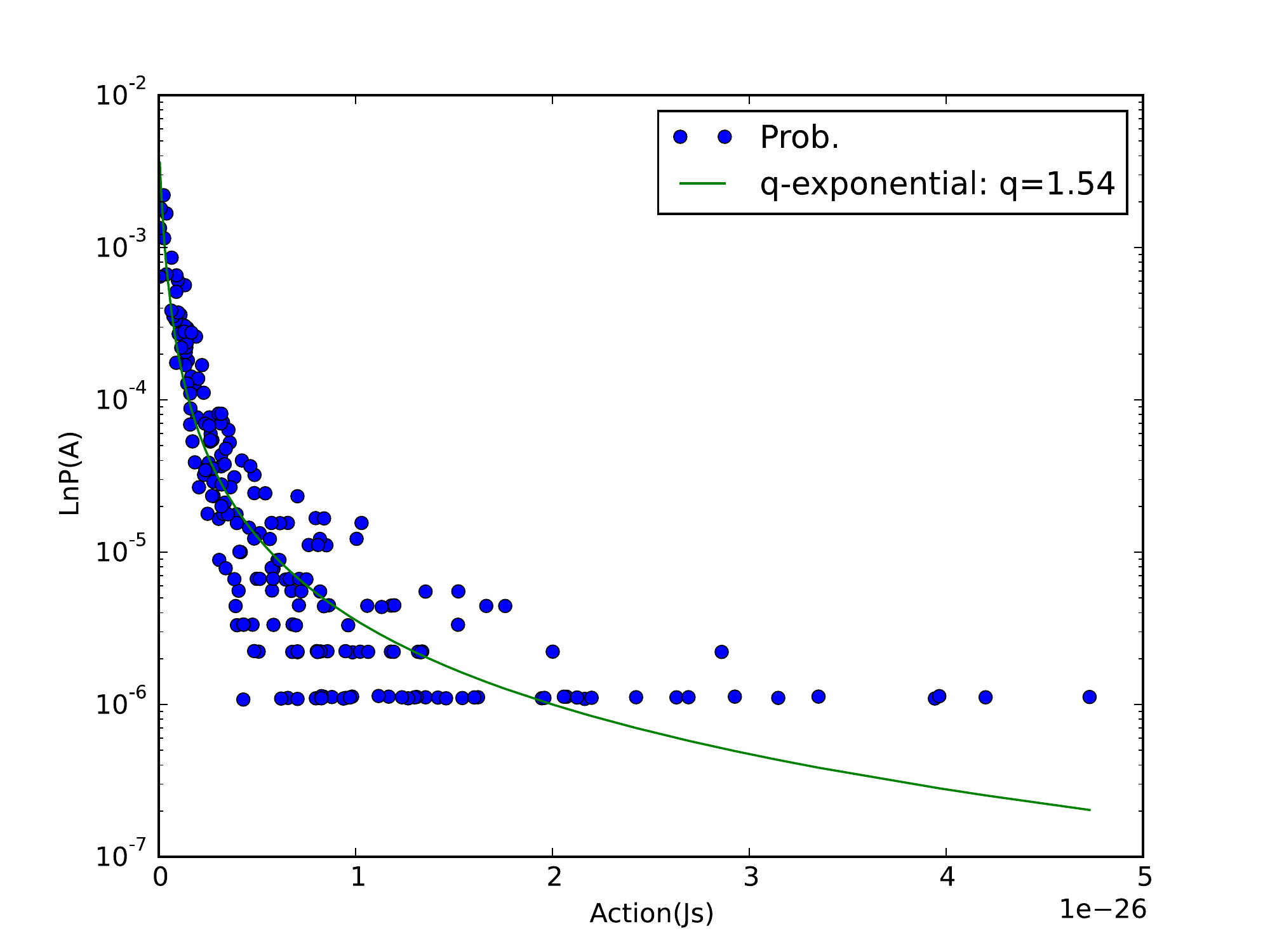}
        \caption{$\alpha=0.5$}
     \end{subfigure}
     \begin{subfigure}[b]{0.43\textwidth}
         \centering
         \includegraphics[width=\textwidth]{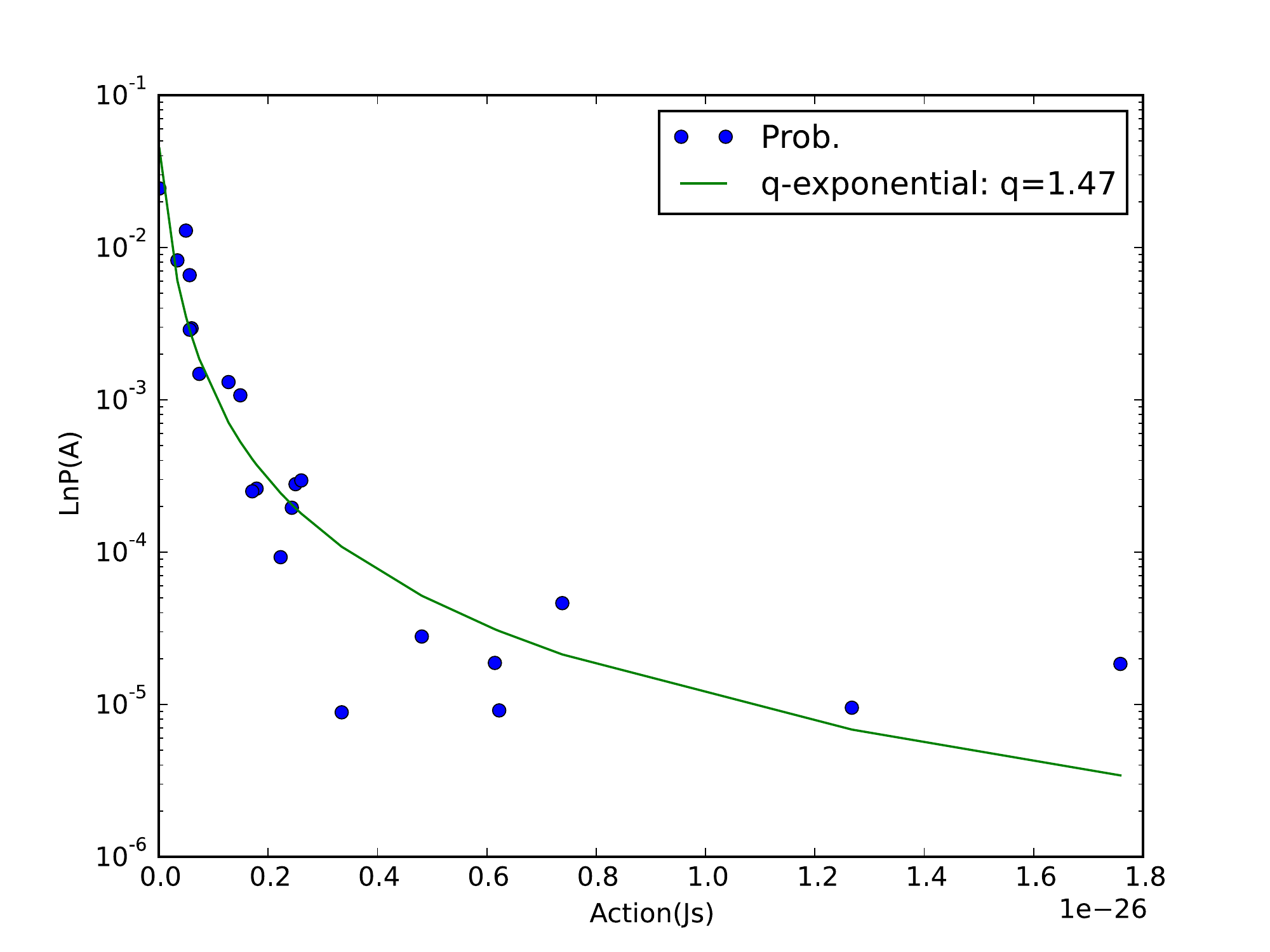}
\caption{$\alpha=0.25$}
     \end{subfigure}
        \caption{Computational simulations for the free particle scenario considering the stable distribution for different values of $\alpha$.}
        \label{fig:liv_stable}
\end{figure}

\begin{figure}[H]
     \centering
     \begin{subfigure}[b]{0.43\textwidth}
         \centering
         \includegraphics[width=\textwidth]{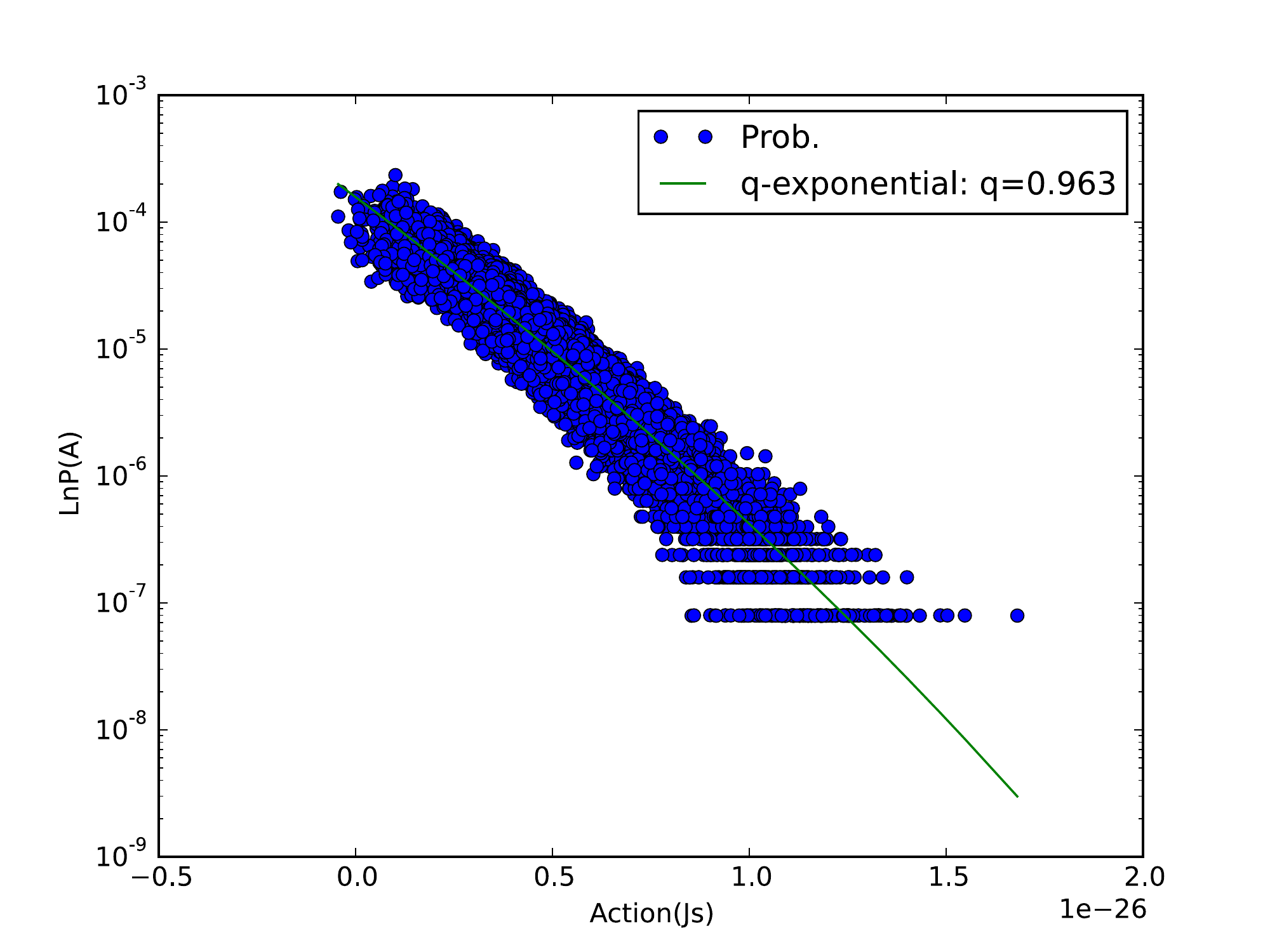}
         \caption{$\alpha=2$}
     \end{subfigure}
     \begin{subfigure}[b]{0.43\textwidth}
         \centering
        \includegraphics[width=\textwidth]{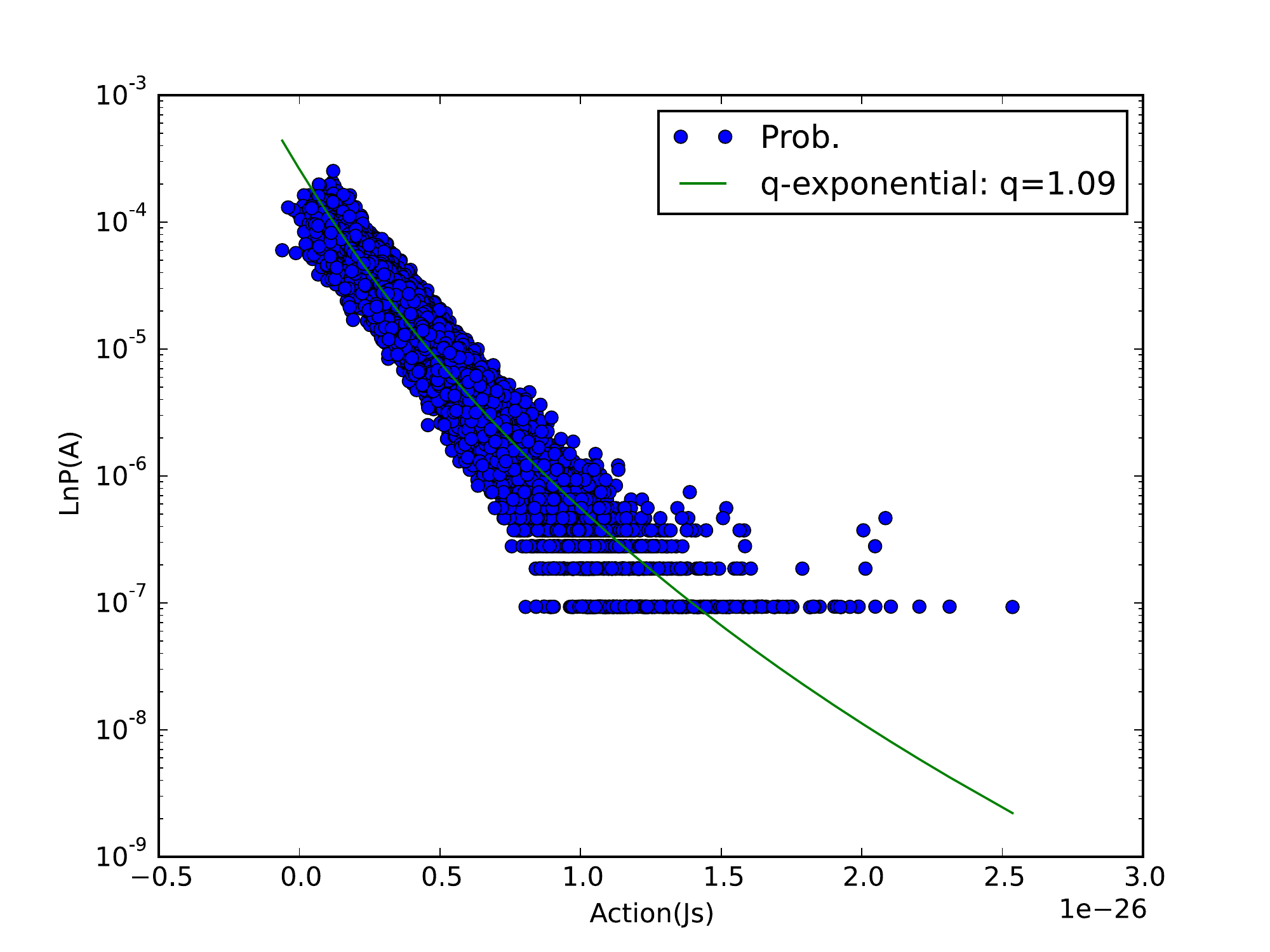}
         \caption{$\alpha=1.75$}
     \end{subfigure}
     \begin{subfigure}[b]{0.43\textwidth}
         \centering
         \includegraphics[width=\textwidth]{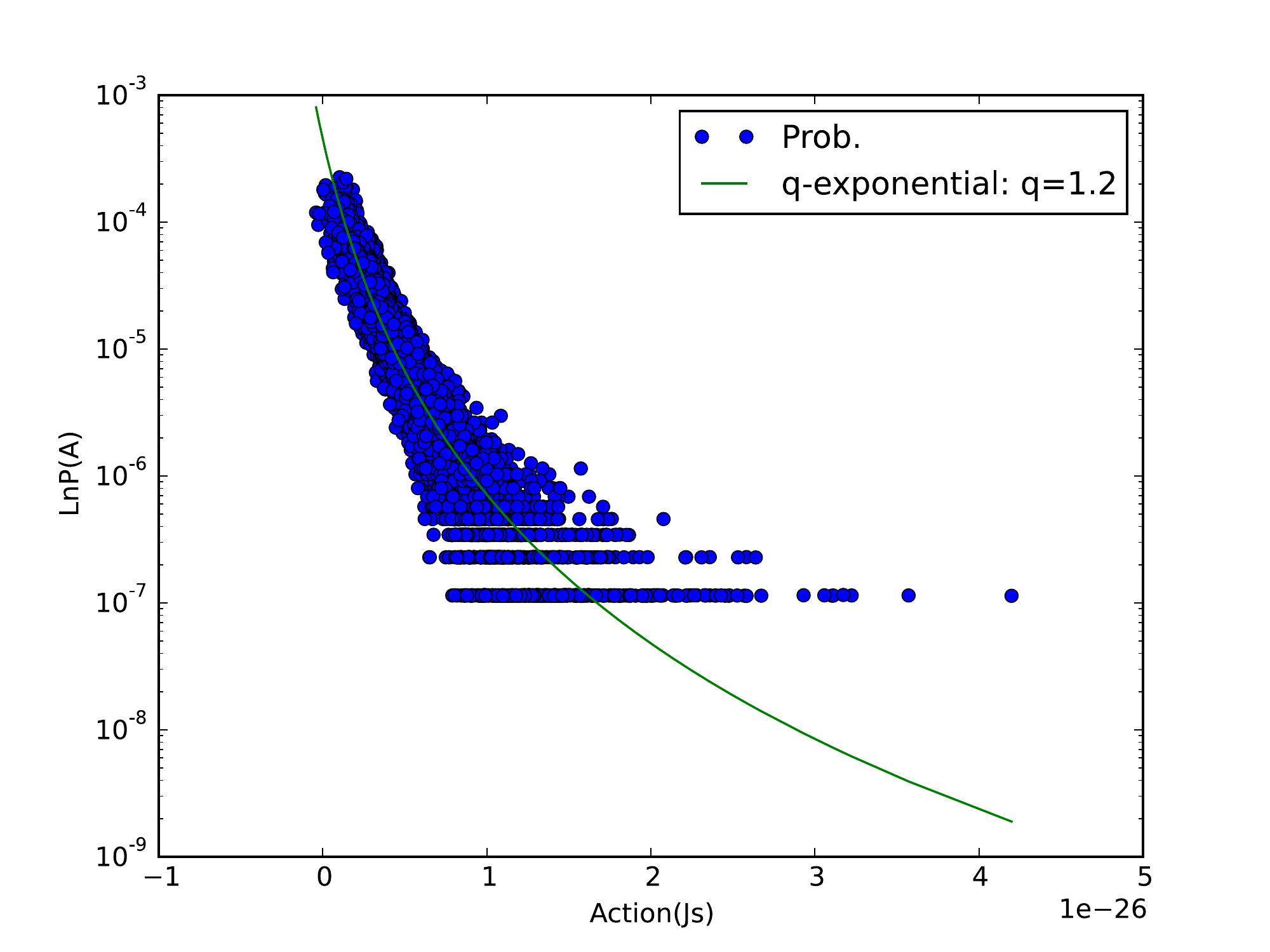}
         \caption{$\alpha=1.5$}
     \end{subfigure}
    \begin{subfigure}[b]{0.43\textwidth}
         \centering
        \includegraphics[width=\textwidth]{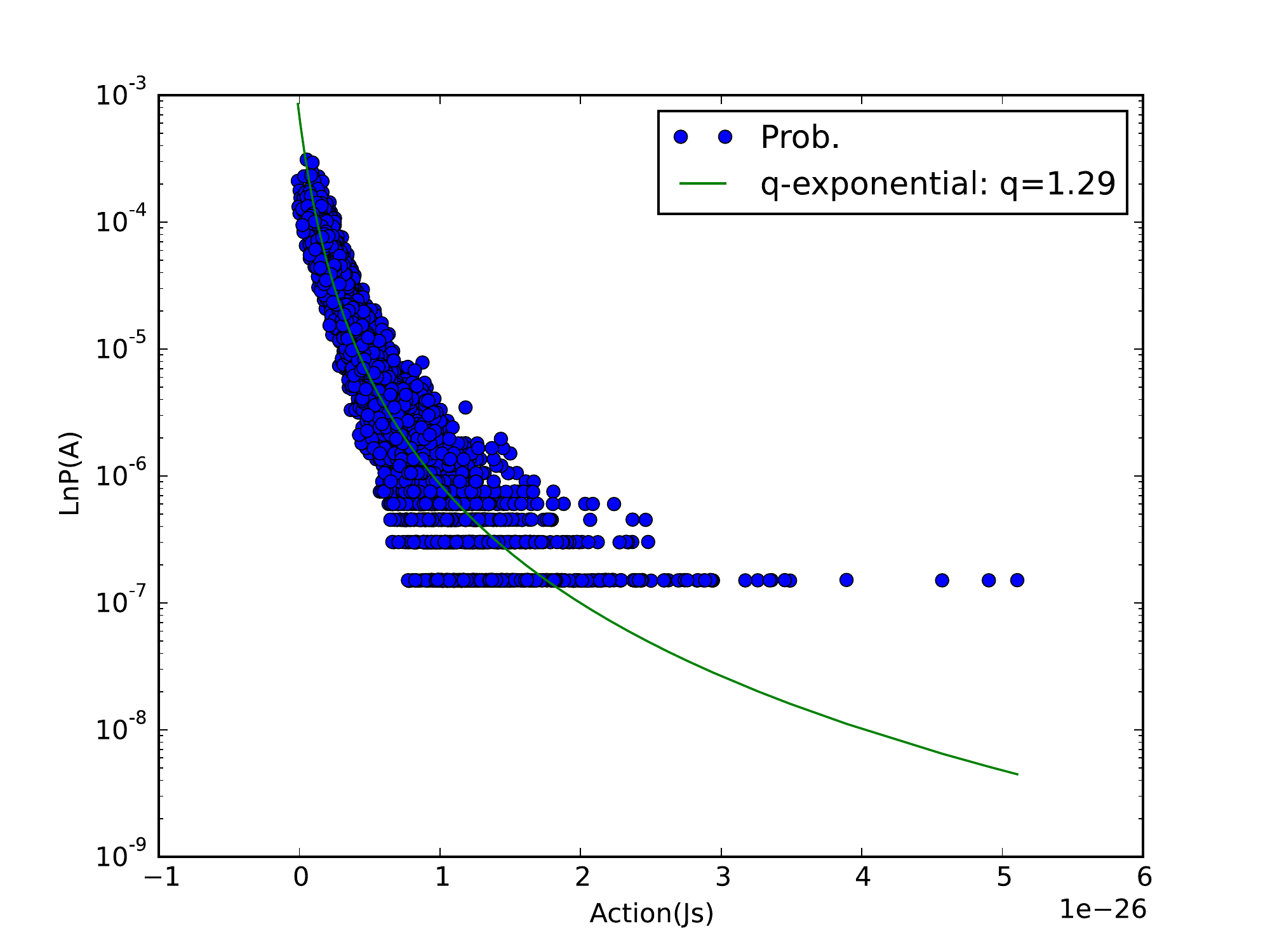}
\caption{$\alpha=1.25$}
     \end{subfigure}
     \begin{subfigure}[b]{0.43\textwidth}
         \centering
         \includegraphics[width=\textwidth]{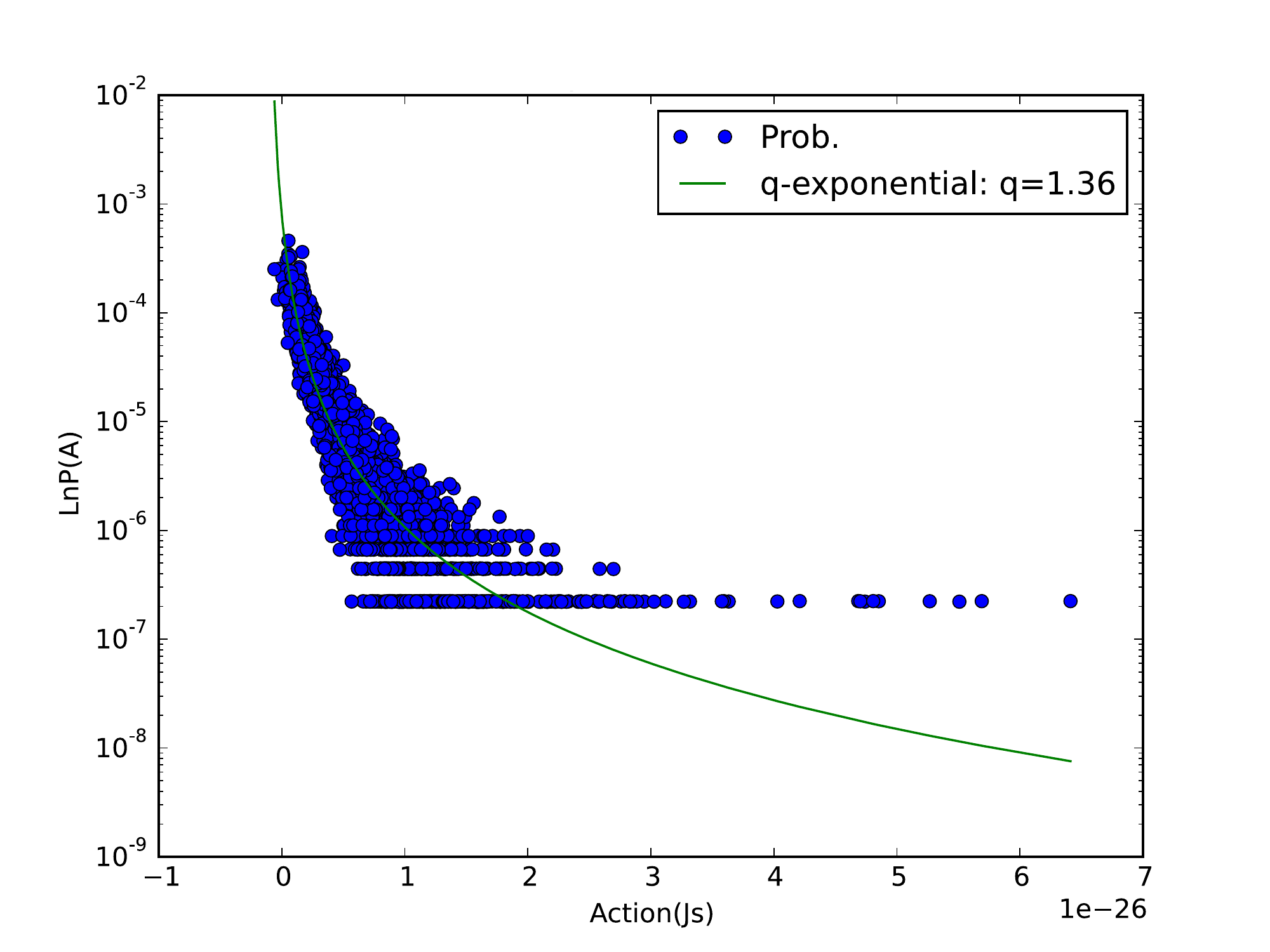}
         \caption{$\alpha=1$}
     \end{subfigure}
     \begin{subfigure}[b]{0.43\textwidth}
        \includegraphics[width=\textwidth]{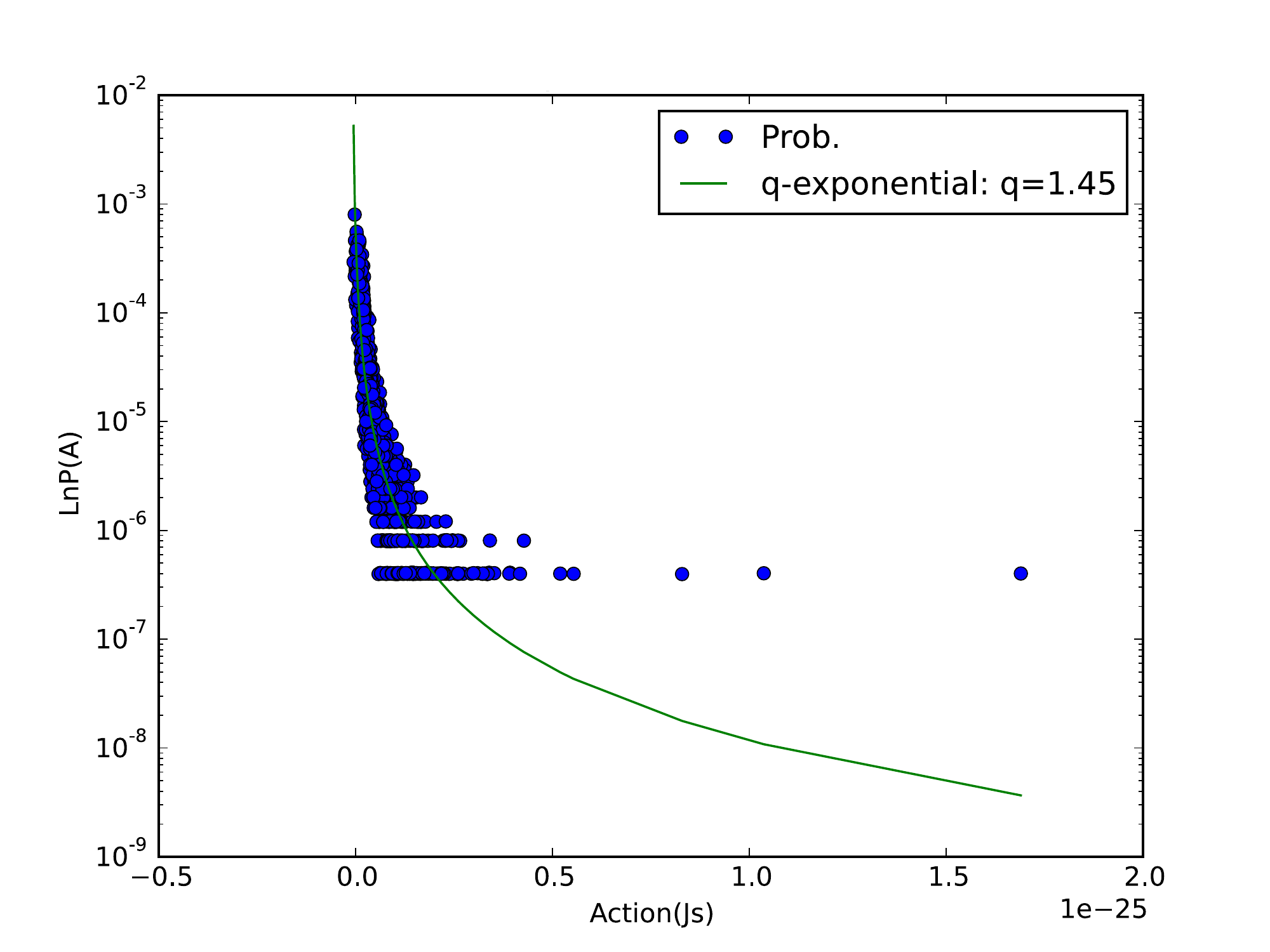}
         \caption{$\alpha=0.75$}
     \end{subfigure}
     \begin{subfigure}[b]{0.43\textwidth}
         \centering
         \includegraphics[width=\textwidth]{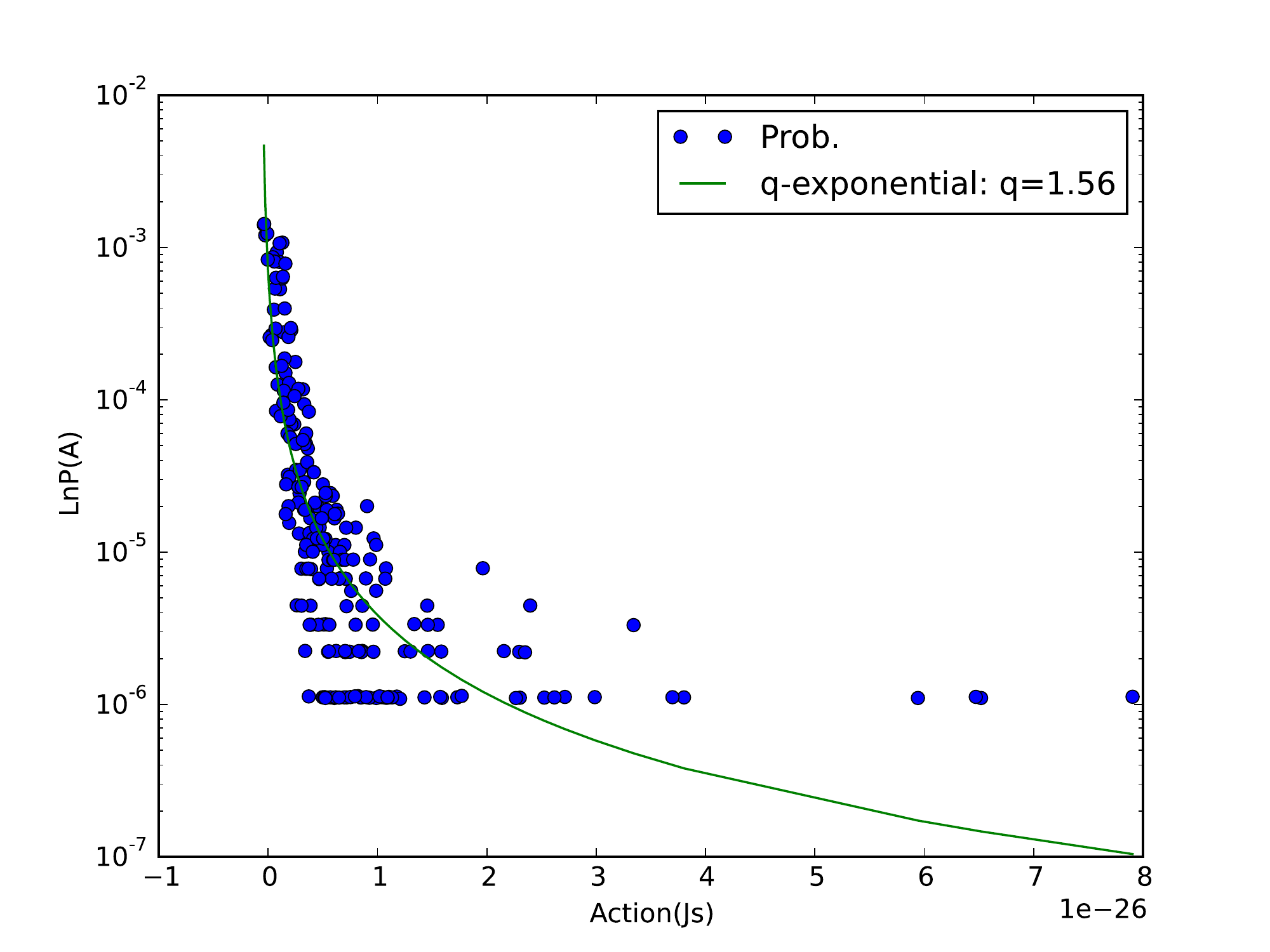}
        \caption{$\alpha=0.5$}
     \end{subfigure}
     \begin{subfigure}[b]{0.43\textwidth}
         \centering
        \includegraphics[width=\textwidth]{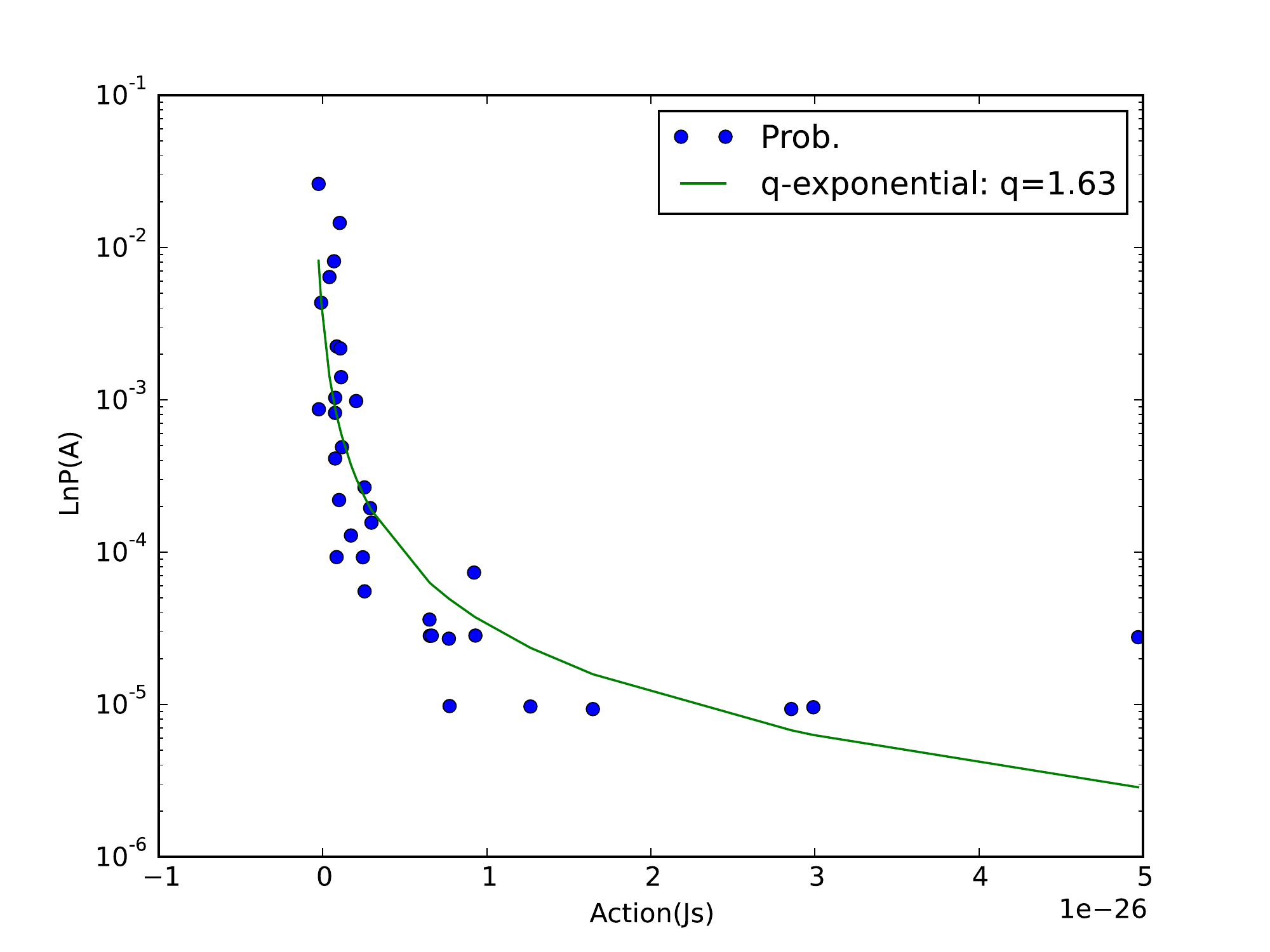}
         \caption{$\alpha=0.25$}
     \end{subfigure}
        \caption{Computational simulations for a particle under a linear potential scenario considering the stable distribution for different values of $\alpha$.}
        \label{fig:const_stable}
\end{figure}

\begin{figure}[H]
     \centering
     \begin{subfigure}[b]{0.43\textwidth}
         \centering
         \includegraphics[width=\textwidth]{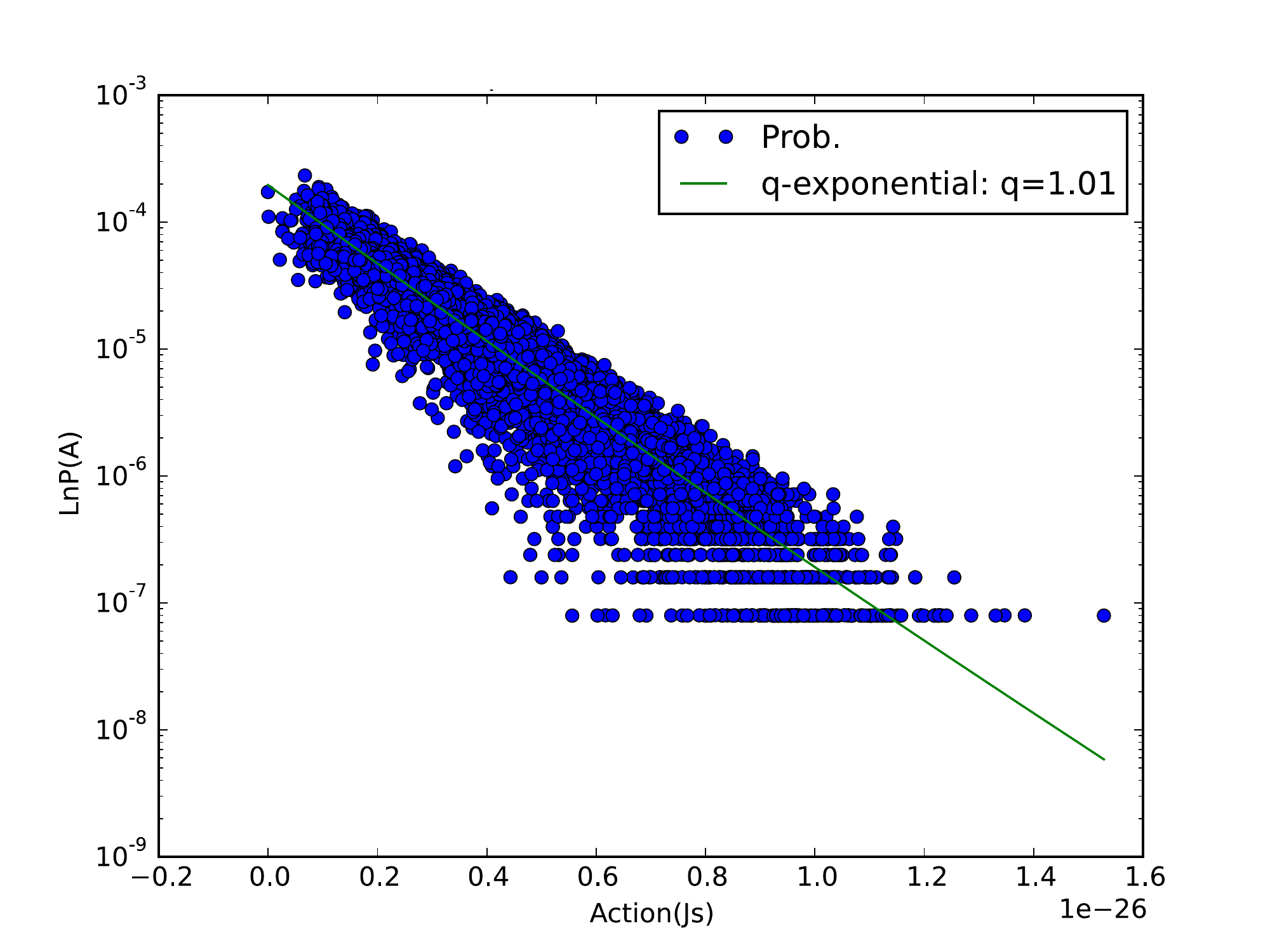}
         \caption{$\alpha=2$}
     \end{subfigure}
     \begin{subfigure}[b]{0.43\textwidth}
         \centering
         \includegraphics[width=\textwidth]{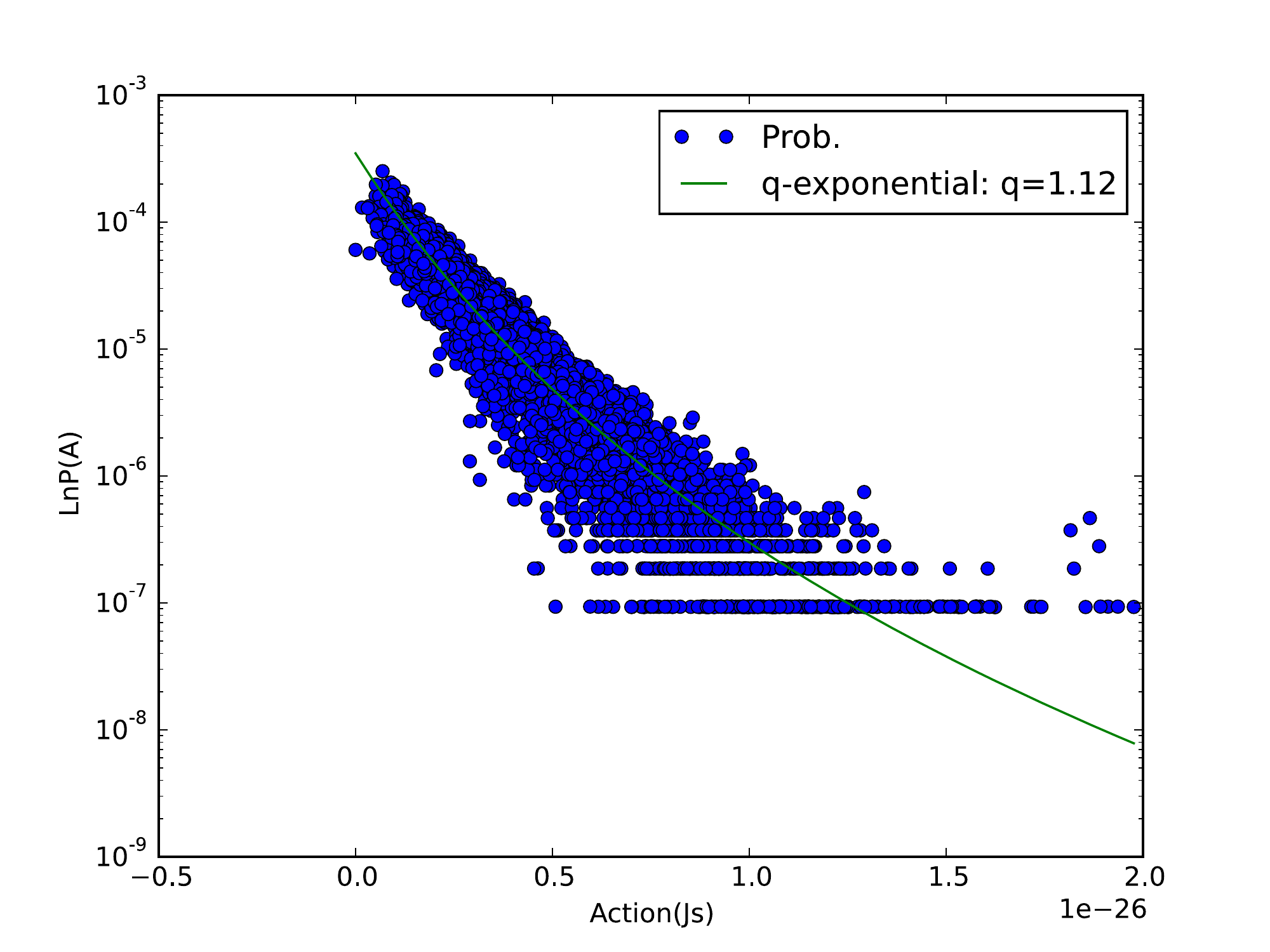}
         \caption{$\alpha=1.75$}
     \end{subfigure}
     \begin{subfigure}[b]{0.43\textwidth}
         \centering
         \includegraphics[width=\textwidth]{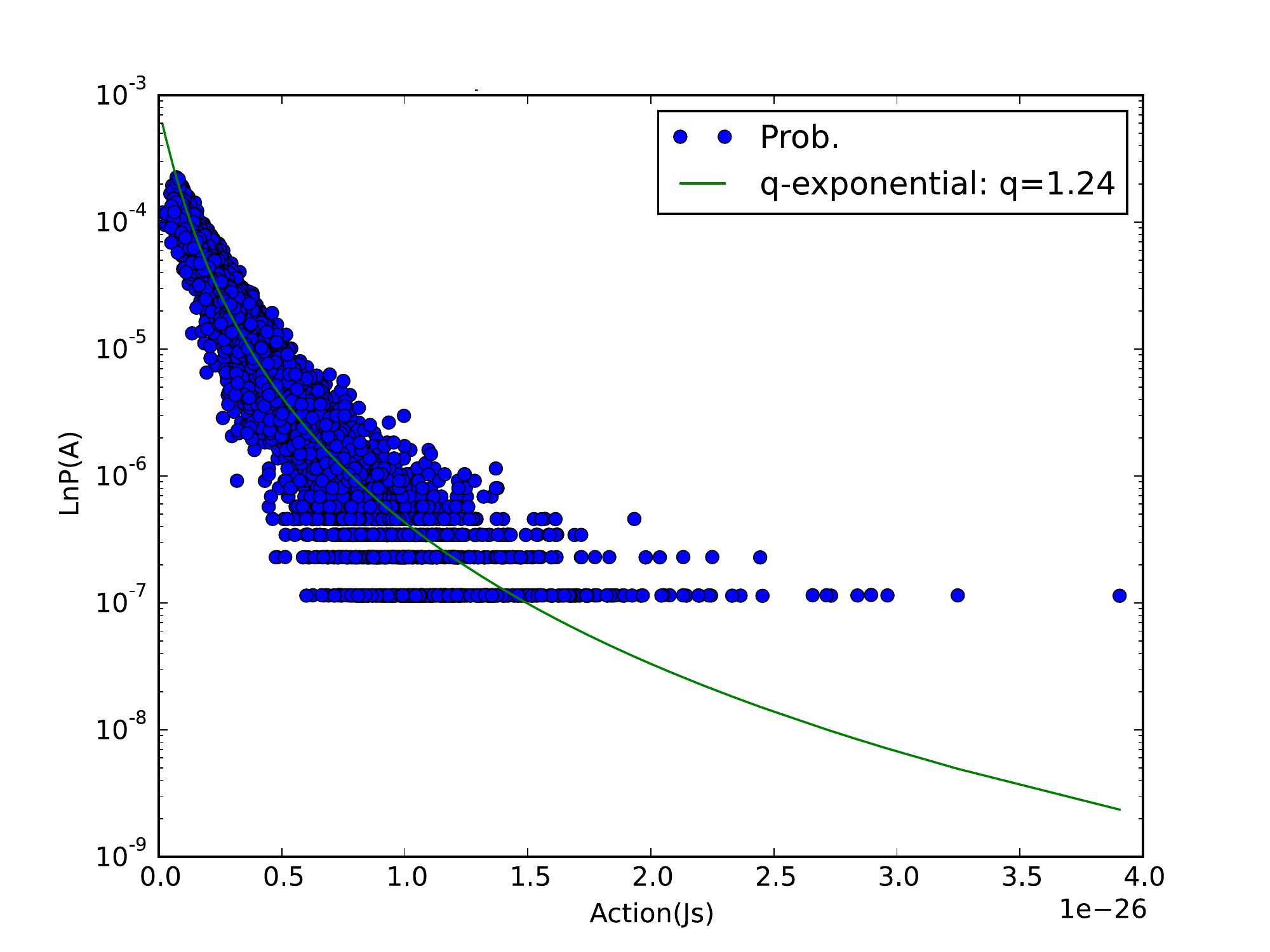}
         \caption{$\alpha=1.5$}
     \end{subfigure}
    \begin{subfigure}[b]{0.43\textwidth}
         \centering
         \includegraphics[width=\textwidth]{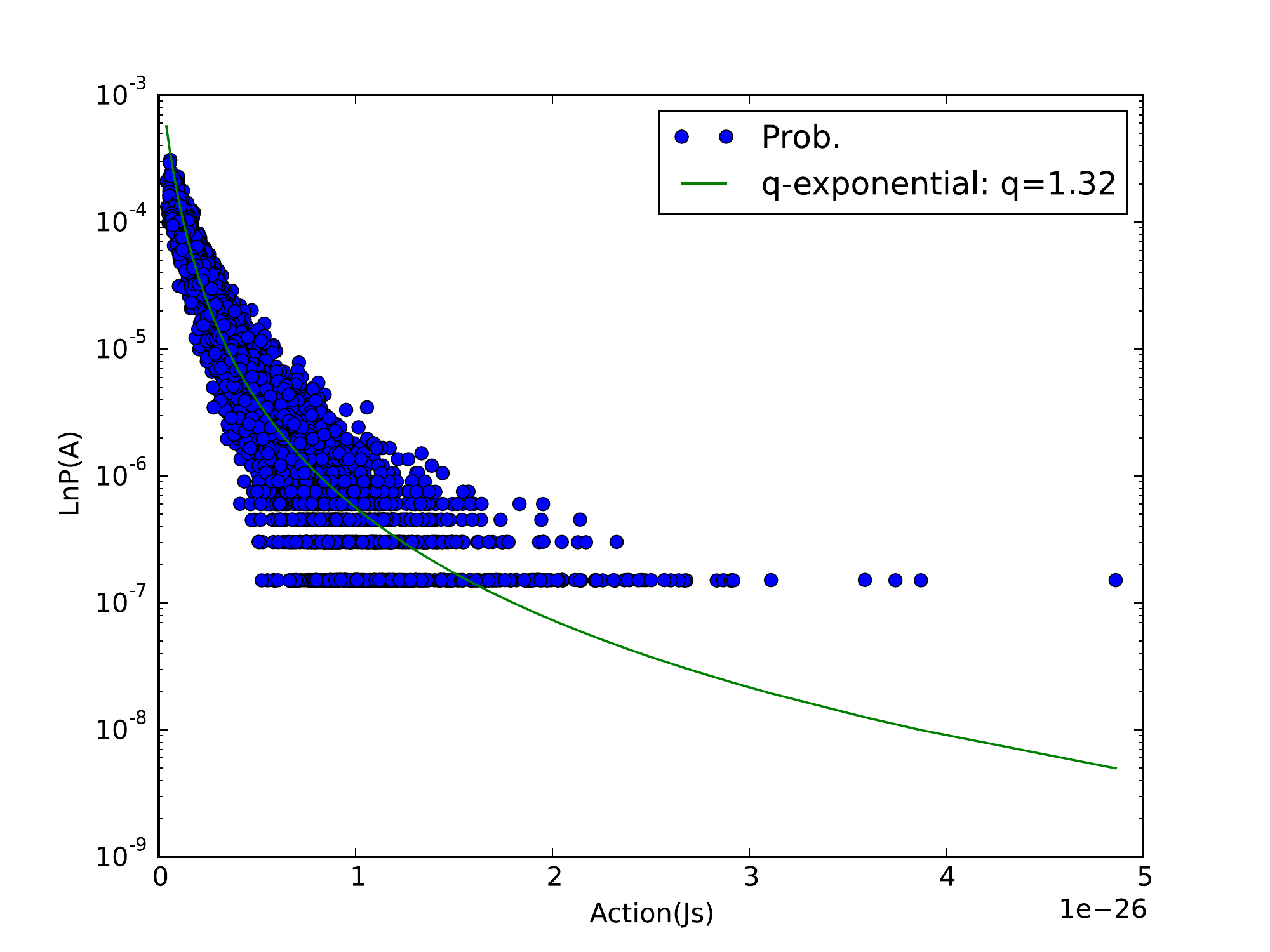}
         \caption{$\alpha=1.25$}
     \end{subfigure}
     \begin{subfigure}[b]{0.43\textwidth}
         \centering
         \includegraphics[width=\textwidth]{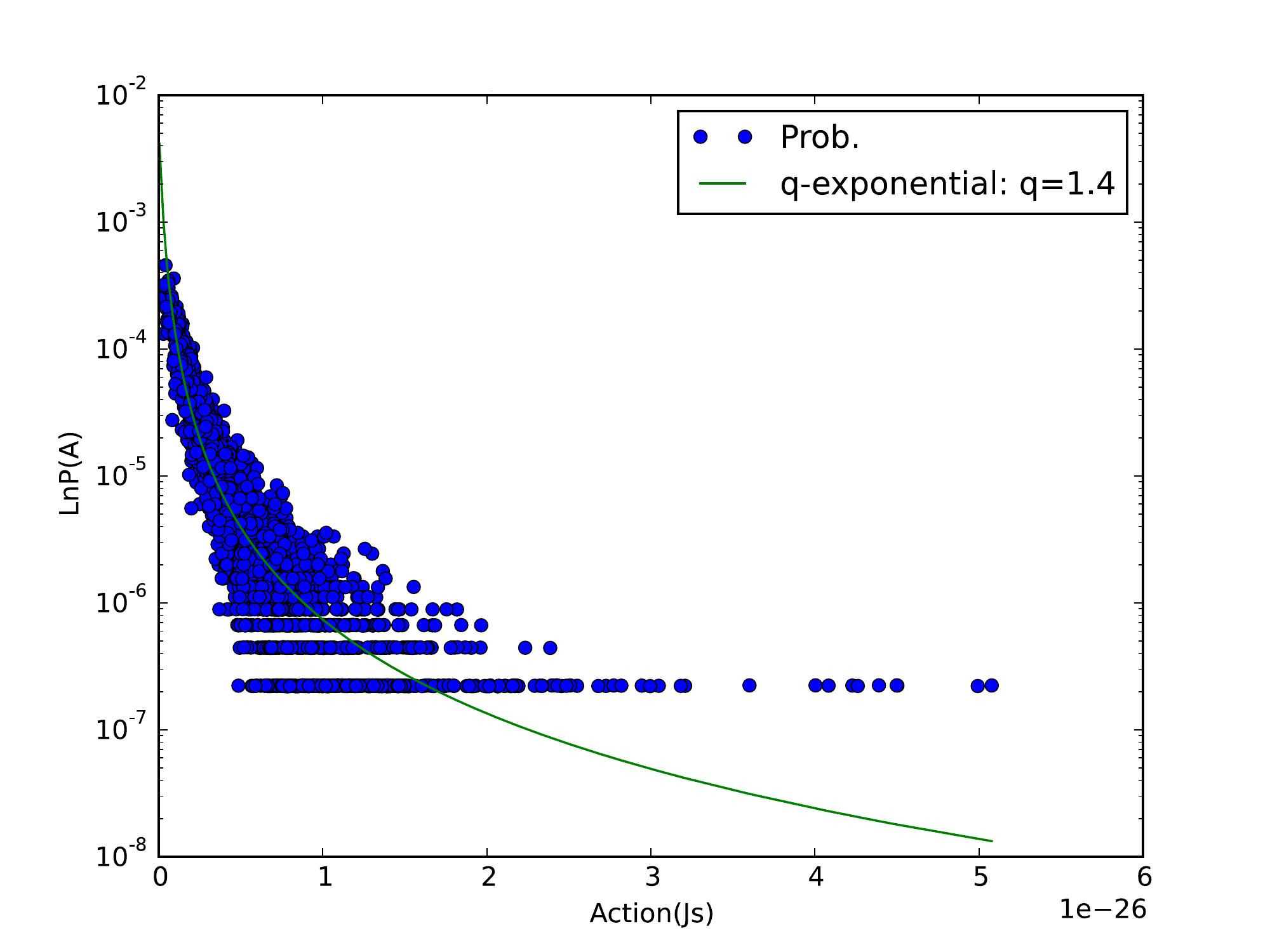}
         \caption{$\alpha=1$}
     \end{subfigure}
     \begin{subfigure}[b]{0.43\textwidth}
        \includegraphics[width=\textwidth]{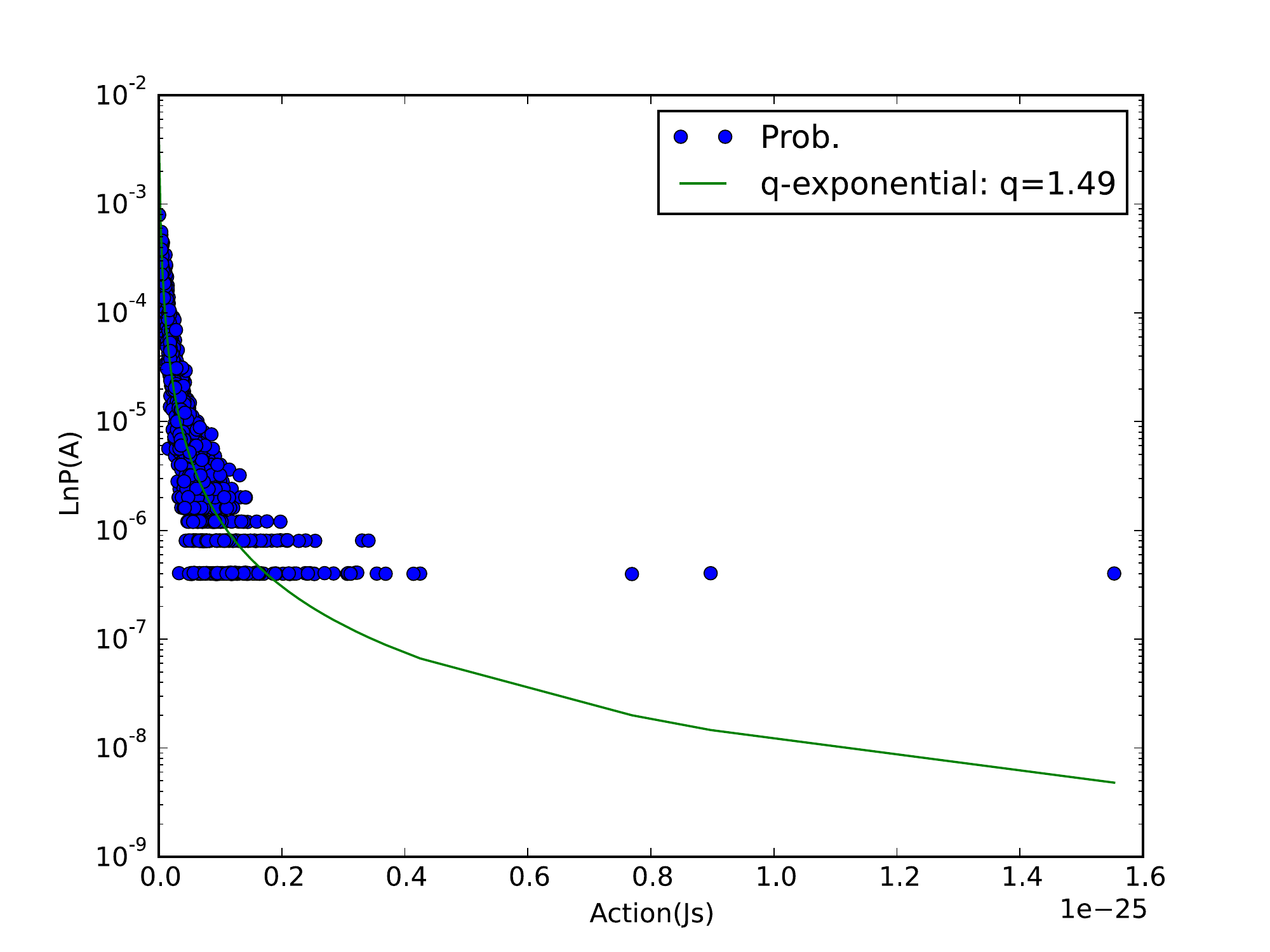}
         \caption{$\alpha=0.75$}
     \end{subfigure}
    \begin{subfigure}[b]{0.43\textwidth}
         \centering
         \includegraphics[width=\textwidth]{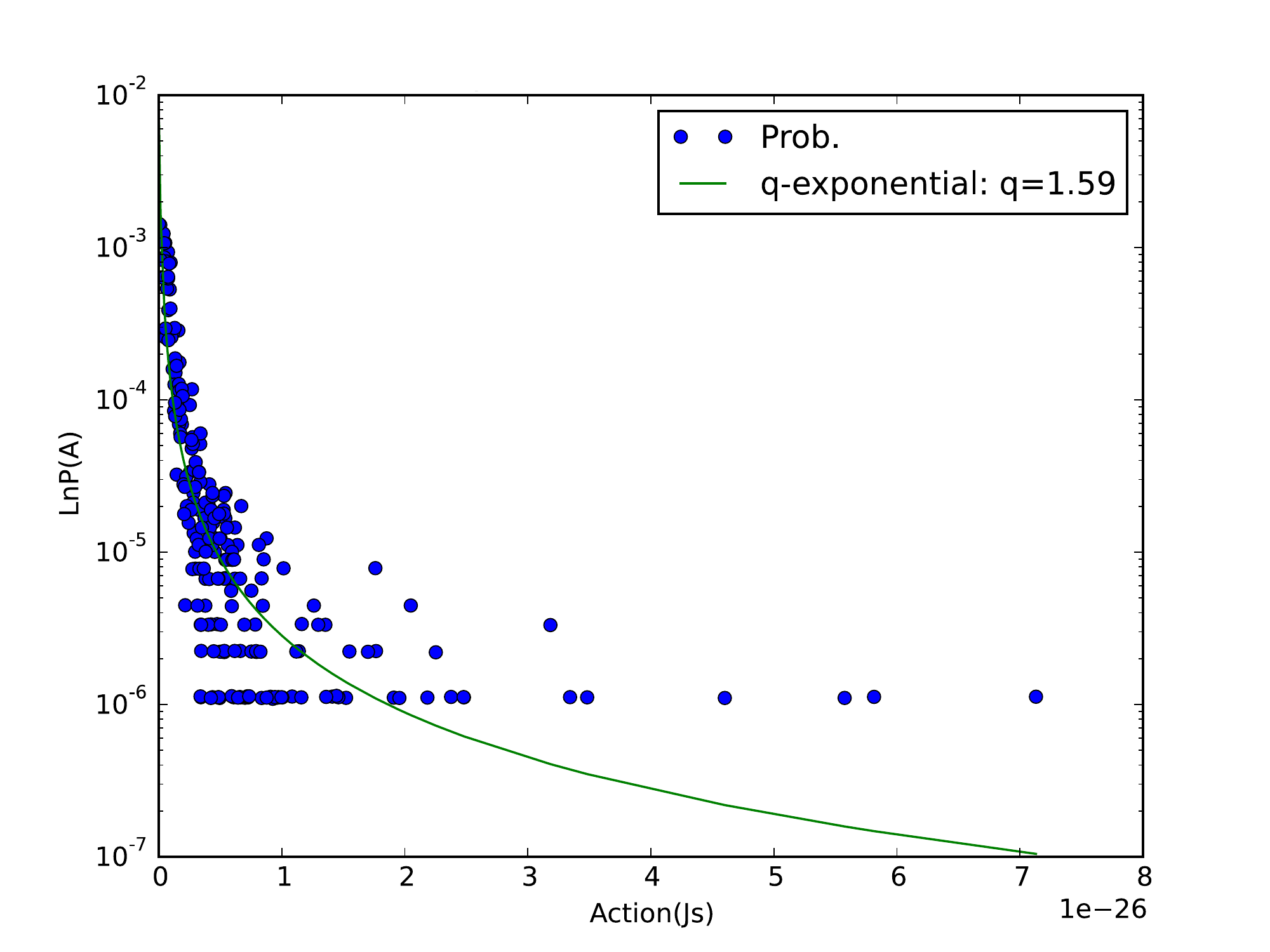}
         \caption{$\alpha=0.5$}
     \end{subfigure}
     \begin{subfigure}[b]{0.43\textwidth}
         \centering
        \includegraphics[width=\textwidth]{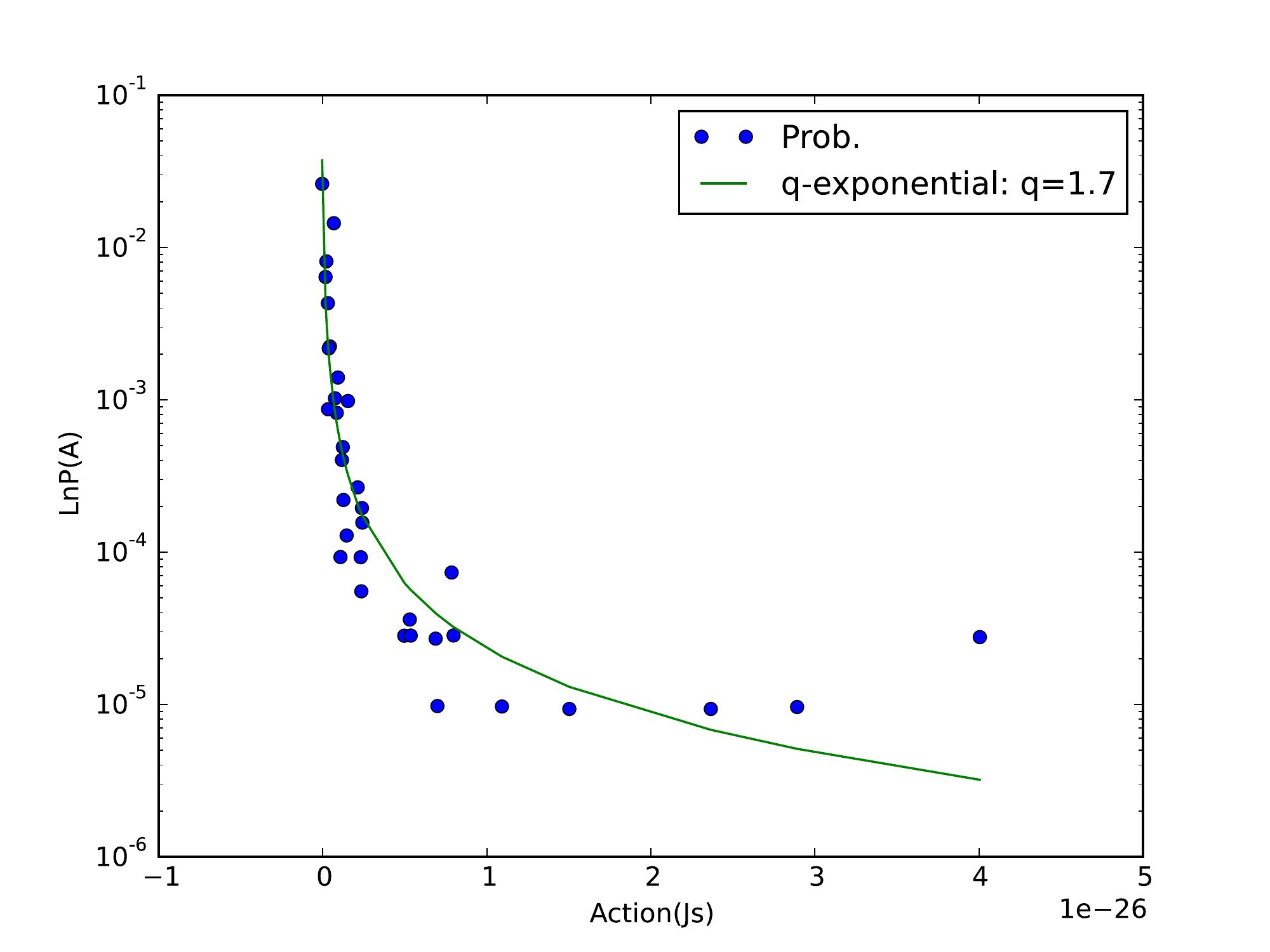}
\caption{$\alpha=0.25$}
     \end{subfigure}
        \caption{Computational simulations for a particle under a quadratic potential scenario considering the stable distribution for different values of $\alpha$.}
        \label{fig:ohs_stable}
\end{figure}

Therefore, once we have rewritten the path probability distribution, now considering the Tsallis entropy, we can perform the computational simulations considering the potentials: null, linear and quadratic, for different values of $\alpha$.
This analysis will allow us to verify the validity of the path probability distribution \eqref{path_gen} as shown below.
In addition, since we are interested on the rare events that lie within the heavy tailed distribution, we have increased the number of paths from  $10^3$ to $10^5$.
The results from the computational simulations for the logarithm of the probability ($\ln P$) versus the action of the paths ($A$) can be seen in the figures \ref{fig:liv_stable}, \ref{fig:const_stable} and \ref{fig:ohs_stable}.

\begin{table}[H]
\centering
\begin{tabular}{|
>{\columncolor[HTML]{FFFFFF}}c |
>{\columncolor[HTML]{EFEFEF}}c |
>{\columncolor[HTML]{FFFFFF}}c |
>{\columncolor[HTML]{EFEFEF}}c |
>{\columncolor[HTML]{FFFFFF}}c |
>{\columncolor[HTML]{EFEFEF}}c |
>{\columncolor[HTML]{FFFFFF}}c |}
\hline
\cellcolor[HTML]{FFFFFF} &
  \multicolumn{2}{c|}{\cellcolor[HTML]{EFEFEF}Null} &
  \multicolumn{2}{c|}{\cellcolor[HTML]{C0C0C0}Linear} &
  \multicolumn{2}{c|}{\cellcolor[HTML]{EFEFEF}Quadratic} \\ \cline{2-7} 
\multirow{-2}{*}{\cellcolor[HTML]{FFFFFF} $\alpha$} & 
  q &
  q-error &
  q &
  q-error &
  q &
  q-error \\ \hline
2    & 0.984 & 0.002 & 0.963 & 0.003 & 1.009 & 0.004 \\ \hline
1.75 & 1.113 & 0.002 & 1.092 & 0.003 & 1.124 & 0.004 \\ \hline
1.5  & 1.227 & 0.003 & 1.203 & 0.004 & 1.236 & 0.004 \\ \hline
1.25 & 1.315 & 0.005 & 1.286 & 0.005 & 1.322 & 0.006 \\ \hline
1    & 1.396 & 0.007 & 1.358 & 0.007 & 1.398 & 0.008 \\ \hline
0.75 & 1.497 & 0.012 & 1.448 & 0.013 & 1.492 & 0.013 \\ \hline
0.5  & 1.536 & 0.027 & 1.558 & 0.035 & 1.593 & 0.032 \\ \hline
0.25 & 1.749 & 0.078 & 1.633 & 0.117 & 1.695 & 0.072 \\ \hline
\end{tabular}
\caption{Values of $q$ and its uncertainty for the null, linear and quadratic potential scenarios and different values of the parameter $\alpha$.}
\label{tab:qtable}
\end{table}

The numerical results of the parameter $q$ are summarized in the table \ref{tab:qtable} for different values of $\alpha$ and the cases of null, linear and quadratic potentials.
The outcomes of our analysis allow us to strongly conclude that the path probability of particles under random motion with stable distributions are, in fact, distributed by the maximization of the Tsallis entropy subjected to the constraints of the stochastic least action principle.
In other words, it means that in order to describe systems with drastic and rare phenomena (named as black swan events), using stable distributions and characterized by the action, the Tsallis entropy is preferred instead of the Shannon entropy, since we are dealing with non-local correlations.

\begin{figure}[H]
    \centering
    \includegraphics[scale=0.5]{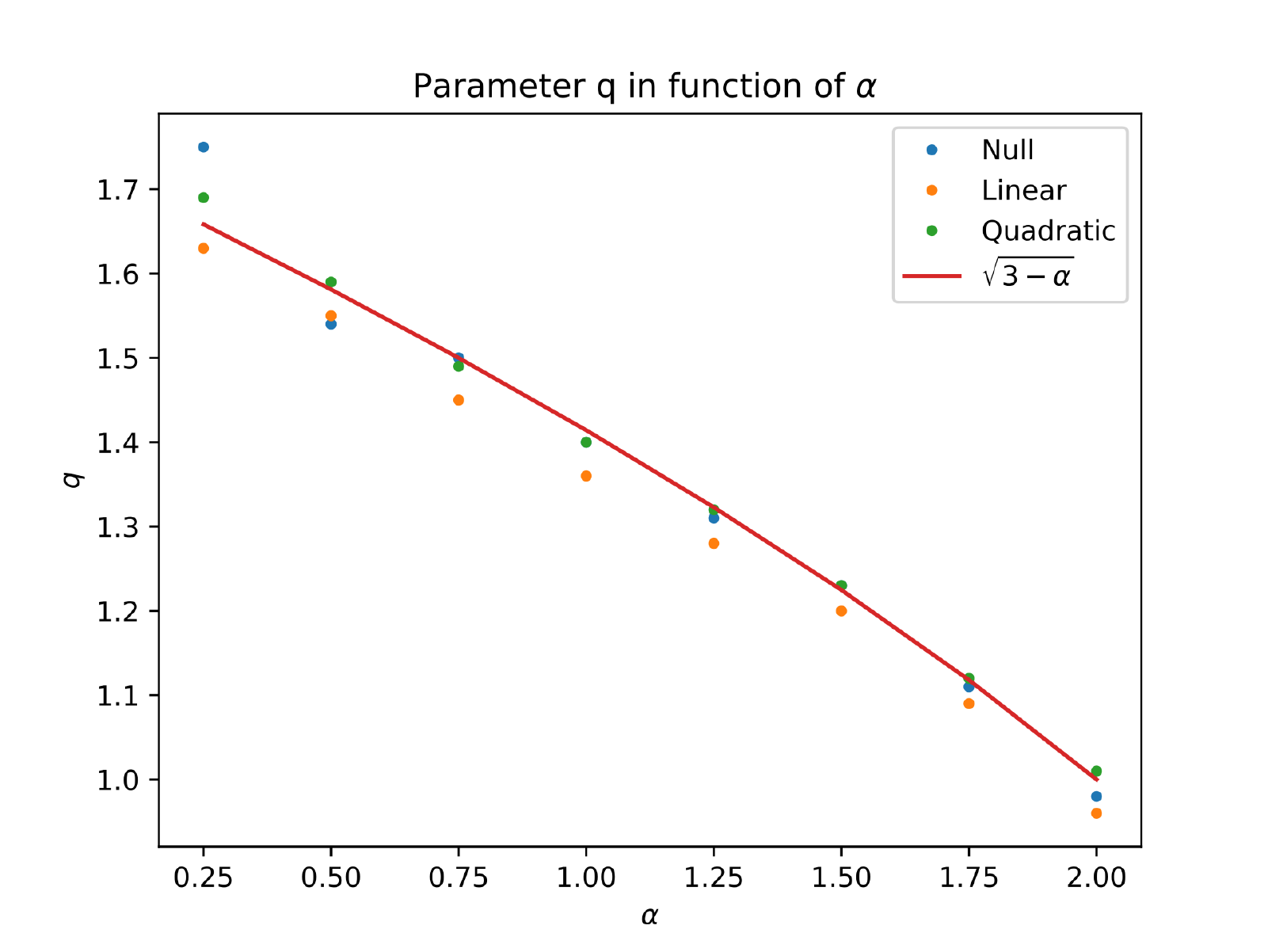}
    \caption{The dots represent the values of $q$ and $\alpha$ to the simulations with different potentials: blue dot for null potential, orange dot for linear potential and green dot for quadratic potential. The continuous line represents the approximated behavior of these two parameters as $q=\sqrt{3 - \alpha}$. }
    \label{fig:qalpha}
\end{figure}

Furthermore, from the results presented in the table \ref{tab:qtable}, it is possible to  observe a relation between the parameters $\alpha$ and $q$, which behaves with good agreement with the data as $q=\sqrt{3 - \alpha}$, see figure \ref{fig:qalpha}.
Actually, this result can be presented as the following statement: the heavier the tails are, stronger is the non-locality effect.
This conclusion is rather expected, since $1-q$ can be understood as a measure of how much the system ``diverges'' from the (normal) behavior, once in the limit $q\rightarrow 1$, the Shannon entropy is recovered.

\section{Conclusions}
\label{sec:conc}

In this work we have generalized the stochastic least action principle to describe the so-called black swan events (of non dissipative systems) with the use of the Tsallis $q$-entropy and heavy tailed distribution.
This approach led to a new path probability distribution, in terms of the $q$-exponential, for the random motion of particles.
Computational simulations have shown the agreement of the present formalism with the path probability of random motion with stable distribution for the scenarios of null, linear and quadratic potential.
These results support the choice of using the Tsallis entropy, since we are dealing with long range interactions of the anomalous diffusion. 
Additionally, the relation between the stable distribution and the Tsallis entropy seems to be well modeled with the form $q=\sqrt{3 - \alpha}$, which can be interpreted as an equivalence between the long range interactions and the occurrence of the black swan events. 
Therefore, we can state that within the SAP formalism, the Tsallis $q$-entropy provides the natural environment for the description of the black swan events.

It is worth recalling that the SAP's main strength consists in prescribing the path probability for different types of energy conserving random motion. Its application to thermodynamic systems can help to shed light on the disparities between the thermodynamics laws and classical mechanics laws \cite{wang1, wang2}. Here, we have only investigated black swan events in energy-conserving dynamical systems. Naturally, a direct extension would be a formulation of black swan events within dissipative systems with the purpose of treating random systems in a wider framework \cite{lin2012, wanganali}. We can mention some scenarios where SAP can be suitably applied to with the technique of path integrals \cite{kleinert}: financial Black-Scholes formalism \cite{black}, models of biological evolution \cite{pease} and the human body neural system \cite{ivancevic}, among others. Efforts in these directions are now under development and will be reported elsewhere.

\subsection*{Acknowledgments}

This study was funded partially by the Coordenação de Aperfeiçoamento de Pessoal de Nível Superior -- Brasil (CAPES) -- Finance Code 001.
R.B. acknowledges partial support from Conselho
Nacional de Desenvolvimento Cient\'ifico e Tecnol\'ogico (CNPq Projects No. 305427/2019-9 and No. 421886/2018-8) and Funda\c{c}\~ao de Amparo \`a Pesquisa do Estado de Minas Gerais (FAPEMIG Project No. APQ-01142-17).



\end{document}